\documentclass[aps,prl,twocolumn,english,superscriptaddress,nolongbibliography]{revtex4-1}
\usepackage{amsmath}
\usepackage{amssymb,scalerel}
\usepackage{graphicx}
\usepackage[colorlinks,bookmarks=false,citecolor=red,linkcolor=blue,urlcolor=blue]{hyperref}
\usepackage{lipsum, color}
\usepackage{wasysym}
\UseRawInputEncoding
\def\Hhat{\hat{H}}

\def\r{{\boldsymbol r}}

\def\i{{\boldsymbol i}}
\def\j{{\boldsymbol j}}
\def\k{{\boldsymbol k}}
\def\q{{\boldsymbol q}}

\def\Q{{\boldsymbol Q}}
\def\q{{\boldsymbol q}}

\newcommand{\be}{\begin{equation}}
\newcommand{\ee}{\end{equation}}
\newcommand{\bea}{\begin{eqnarray}}
\newcommand{\eea}{\end{eqnarray}}
\newcommand{\beas}{\begin{eqnarray*}}
\newcommand{\eeas}{\end{eqnarray*}}
\newcommand{\ve}[1]{\boldsymbol{#1}}

\begin{document}
\title{
Spin chain on a metallic surface: Dissipation-induced order vs. Kondo  entanglement
}
\author{Bimla Danu}
\affiliation{Institut f\"ur Theoretische Physik und Astrophysik and W\"urzburg-Dresden Cluster of Excellence ct.qmat, Universit\"at W\"urzburg, 97074 W\"urzburg, Germany}
\author{Matthias Vojta}
\affiliation{Institut f\"ur Theoretische Physik and W\"urzburg-Dresden Cluster of Excellence ct.qmat, Technische Universit\"at Dresden, 01062 Dresden, Germany}
\author{Tarun Grover}
\affiliation{Department of Physics, University of California at San Diego, La Jolla, CA 92093, USA}
\author{Fakher F. Assaad}
\affiliation{Institut f\"ur Theoretische Physik und Astrophysik and W\"urzburg-Dresden Cluster of Excellence ct.qmat, Universit\"at W\"urzburg, 97074 W\"urzburg, Germany}

\date{\today}


\begin{abstract}
We explore the physics of  a spin-1/2  Heisenberg chain with  Kondo interaction, $J_k$, to a two-dimensional  electron  gas.
At weak  $J_k$ the  problem  maps  onto  a  Heisenberg  chain locally coupled to  a dissipative Ohmic bath.  At the decoupled fixed
point, the dissipation is a marginally relevant perturbation and drives long-range antiferromagnetic order along the chain.
In the  dynamical spin structure factor we observe a quadratic low-energy dispersion akin to Landau-damped Goldstone modes.
At large  $J_k$ Kondo screening  dominates,  and the spin correlations of the chain inherit the power law of the host metal, akin to a paramagnetic heavy Fermi liquid.
In both phases we  observe heavy  bands near the  Fermi energy  in the  composite-fermion spectral  function.
Our  results, obtained from auxiliary-field quantum Monte Carlo simulations, provide a unique negative-sign-free realization of a quantum transition between an antiferromagnetic metal and a heavy-fermion metal.  We discuss  the relevance of our  results in the context of scanning tunneling spectroscopy experiments of magnetic adatom chains on metallic surfaces.

\end{abstract}
\maketitle


%
\textit{Introduction.}
A spin-1/2 antiferromagnetic chain  embedded in a higher-dimensional metal, with  Kondo  coupling $J_k$ between spins and electrons, represents an arena for rich physics. For  two-dimensional metals, this relates to scanning tunneling microscopy (STM)
experiments, with the  ability to  build and probe assemblies of  magnetic  adatoms on  surfaces~\cite{Toskovic2016, Choi2017, Moro2019, ChoiRMP2019, Danu2019}.  In higher dimensions,  Yb$_2$Pt$_2$Pb  provides  a realization of  one-dimensional spin chains  embedded  in a  three-dimensional metal \cite{Wu16,Gannon19}.   Due to the dimensionality  mismatch, such a system remains metallic even for a half-filled conduction band. It  can host a  variety of phases  that   include  Kondo-breakdown or  orbital-selective Mott states~\cite{Vojta10,Danu2020},
heavy-fermion  physics in which the magnetic spins,  albeit sub-extensive,  participate in the Luttinger volume, as  well as non-Fermi-liquid states~\cite{Classen18}.  The  understanding of quantum transitions between these states is of considerable interest  both  experimentally and  theoretically.

\begin{figure}[htbp]
\includegraphics[width=0.48\textwidth]{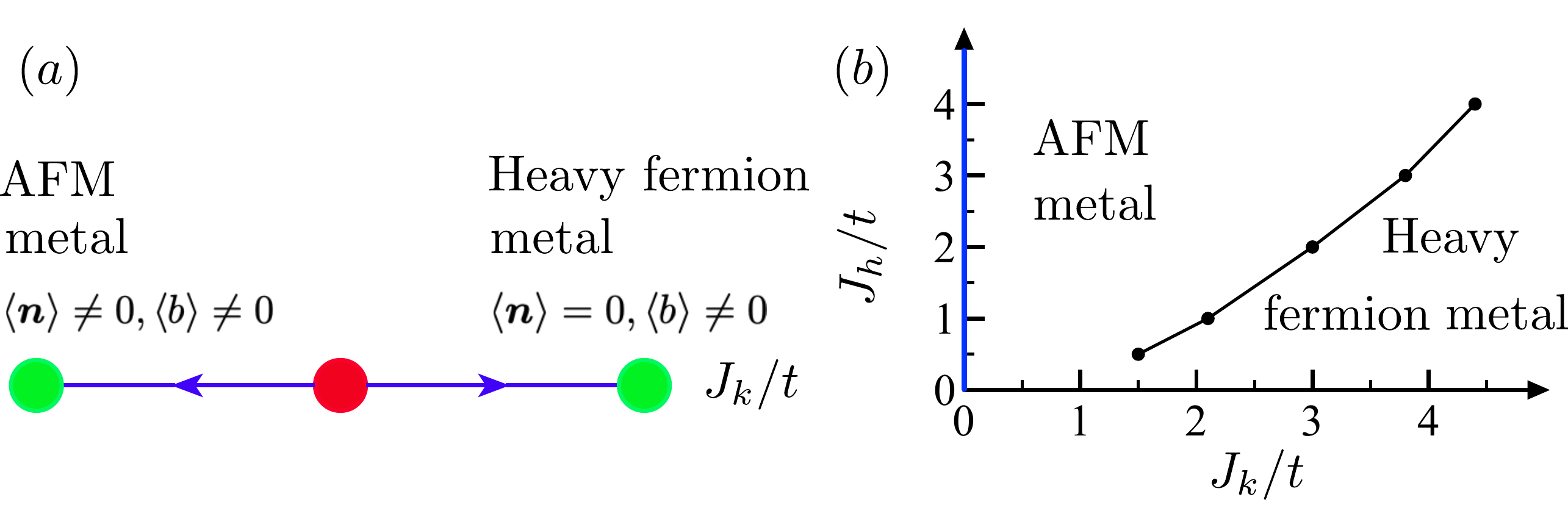}
\caption{(a)  RG flow diagram  as suggested by the  QMC  data.   Green   (red)  bullets  correspond  to
 phases (critical points).  We observe  an   antiferromagnetic order-disorder
transition  with $\langle  \ve{n} \rangle  $  the  O(3)  order parameter.    In both phases Kondo
screening,  corresponding to a  Higgs condensate  $\langle b \rangle \ne 0$,  is  present.
 (b) Phase diagram  in the  $J_h$ versus  $J_k$ plane  as extracted from  QMC  simulations
 at $\beta \propto L^z$  with $z=2$.  The blue line at $J_k=0$ represents the
  decoupled Heisenberg chain  that is unstable to  dissipation-induced ordering  upon  Kondo coupling
  to the fermions.}
\label{phsgm_RG_QMC}
\end{figure}

In this  letter, we will  consider  the  above  setup   for  two-dimensional electrons in the presence of a Fermi surface.
In the limit of weak Kondo coupling,   one  can  follow  the  Hertz-Millis  approach~\cite{AHertzPRB1976, AJMillis1993} and perturbatively
integrate out  the  fermions to arrive at an effective description of the spin chain  locally coupled   to an Ohmic  bath~\cite{Pwerner2005, WernerPRL2005, CazalillaPRL2006, Zheng2018, Weber2021}.
As argued in Ref.~\onlinecite{Weber2021}, for an O(3) quantum rotor model coupled to an Ohmic bath,  the  dissipation is marginally  relevant   and  leads to long-range magnetic ordering along the chain.  Hence, unlike in conventional heavy-fermions systems where Ruderman-Kittel-Kasuya-Yosida (RKKY) interactions directly drive magnetic ordering~\cite{RudermanKittel1954, KYosida1957}, here the ordering is stabilized only by the dissipation. As the Kondo coupling increases, Kondo screening will compete with dissipation-induced  ordering.    In  particular, in the  strong-coupling limit, it is expected that the spin-rotation symmetry will be restored in the chain, and the  spin-spin  correlations of the chain will inherit the  power-law  decay of the host metal.
The  physics of the Heisenberg spin chain on a metallic  surface can  hence  be  cast  into the flow  diagram of  Fig.~\ref{phsgm_RG_QMC}(a)  where  Kondo-singlet formation  and dissipation-induced order compete.

\noindent \textit{Model  and   Method. }
Our starting point is the Hamiltonian for a spin-1/2 chain on a metallic surface,
\begin{eqnarray}
 \Hhat &= &-t\sum_{\langle \i,\j\rangle}\big(\hat{\ve{c}}^{\dagger}_{\i} \hat{\ve{c}}^{}_{\j}+\text{H.c}\big)
 +\frac{J_k}{2}\sum^L_{\ve{r} =1}  \hat{\ve{c}}^{\dagger}_{\ve{r}}  \ve{\sigma} \hat{\ve{c}}^{}_{\ve{r}}
 \cdot \hat{\ve{S}}_{\ve{r}}\nonumber\\&&+J_h\sum^{L}_{\ve{r}=1} \hat{\ve{S}}_{\ve{r}} \cdot \hat{\ve{S}}_{\ve{r}+\Delta\ve{r}}.
\label{model_ham}
\end{eqnarray}
Here, the summation $\sum_{\langle \i,\j\rangle}$
runs over nearest neighbors of a square-lattice, $L \times L$,  conducting substrate,  $t$ is the hopping matrix element,  and $\hat{\ve{c}}^{\dagger}_{\ve{i}} = \big(\hat{c}^{\dagger}_{\ve{i},\uparrow}, \hat{c}^{\dagger}_{\ve{i},\downarrow}\big) $
is a spinor  where  $ \hat{c}^{\dagger}_{\ve{i},\uparrow (\downarrow)} $ creates an electron at site $\ve{i}$ with $z$-component of spin
$1/2$ ($-1/2$). $J_k$  is the antiferromagnetic Kondo coupling between  spins  and  conduction electrons, $J_h$ is the
 antiferromagnetic Heisenberg coupling, $\ve{\hat{S}}_{\ve{r}}$ are  spin-1/2 operators and  $L$ is the length of the chain.
We consider an array of ad-atoms at  an interatomic spacing $\Delta \ve{r} =  (a,0)$ with $a=1$  and  periodic boundary conditions are used along the spin chain as well as  for  the conduction electrons.  Translation by  $(a,0)$   is a symmetry of the problem  such that crystal  momentum $\ve{k}$ along  the  chain is  conserved up to a reciprocal lattice vector.  This  model, including the Heisenberg exchange, is   motivated  by the STM work  of Ref.~\onlinecite{Toskovic2016}.

In the absence of Kondo coupling and at $J_h \neq 0$, the local moments at low-energies are described by a Luttinger liquid action ${\mathcal S}_{\text{chain}}$. Denoting the fluctuating antiferromagnetic (AFM) order parameter as $\ve n$, in this theory $\langle \ve{n}(\ve{r},\tau) \cdot \ve{n}(\ve{0},0) \rangle \sim \sqrt{\log(r^2 + \tau^2)} / \sqrt{r^2+\tau^2} $.   Here $r= |\ve{r}|$.  When $J_k \neq 0$, one may proceed by integrating out  the conduction electrons  and  obtain an action up to  second order in  $J_k$ as   $ {\mathcal S} = {\mathcal S}_{\text{chain}} +  {\mathcal S}_{\text{diss}}(\ve n)   $  with
\begin{equation}\label{action_ef}
{\mathcal S}_{\text{diss} }(\ve n)  = \frac{J^2_k}{8} \int d\tau d\tau^\prime \sum_{\ve{r},\ve{r}'} {\ve{n}}_{\ve{r}} (\tau)  \chi^0 (\ve{r}-\ve{r}',  \tau-\tau^\prime)   {\ve{n}}_{\ve{r}'}(\tau^\prime).
\end{equation}
where  $\chi^0$  is  the  antiferromagnetic spin susceptibility of the conduction electrons  and  ${\mathcal S}_{\text{chain}} $  the  action of the  spin chain.   For  generic, non-nested two-dimensional  electrons at finite density, $\chi^{0}(\ve{r} = 0, \tau) \sim 1/\tau^2$, while  $\chi^{0}(\ve{r}, \tau = 0) \sim 1/r^3$. Using power counting, one observes that while the long-range $1/r^3$ spatial decay of $\chi^{0}$ is irrelevant at the $J_k=0$ fixed point, the long-range $1/\tau^2$ decay in the time direction is not innocuous, and at the leading order, corresponds to an dissipative Ohmic bath that is marginal in the renormalization-group sense. In fact, as argued in~\cite{Weber2021}, such a dissipative coupling is a marginally  relevant operator  that triggers long-range order.  To  avoid the  negative-sign-problem  we employ a particle-hole-symmetric conduction band such  that the  Fermi  surface is nested.  As  shown in supplemental material~\cite{suppl} this leads  to a multiplicative logarithmic correction to  $\chi^{0}$: $\chi^{0}(\ve{0}, \tau) \sim \log^2(\tau)/\tau^2$. Therefore, at small $J_k$, the logarithmic enhancement only increases the tendency for the system to become ordered due to dissipation.   A  particularity of the  nested  Fermi  surface  is a  directional  dependence of $\chi^{0}(\ve{r} , 0)$~\cite{suppl}. For a  chain along the $(a,0)$     direction  $\chi^{0}(\ve{r} , 0) \sim 1/r^4$.
At  $J_k \gg J_h, t$ the local moments prefer to form local singlets with the conduction electrons, thereby  resulting in a paramagnetic phase.

We  simulate the Hamiltonian  of  Eq.~(\ref{model_ham}) using the  auxiliary-field  quantum Monte Carlo (AFQMC)~\cite{Blankenbecler81, White89} implementation of the  Algorithms for Lattice Fermions (ALF)~\cite{ALF_v1, ALF_v2}  library.   The model   falls in  the  general  category of  spin-fermion Hamiltonians~\cite{SatoT17_1}   that  do  not  suffer from the sign  problem.    Our simulations are  based on  the finite-temperature grand-canonical AFQMC~\cite{Assaad08_rev, Capponi2001}.  To  reduce  finite-size  effects  we  have   included  an orbital  magnetic  field  of  magnitude   $B = \Phi_0 /L^2 $    where  $\Phi_0$ is the  flux  quantum~\cite{Assaad01}.

\begin{figure}[htbp]
\centering
\includegraphics[width=0.2345\textwidth]{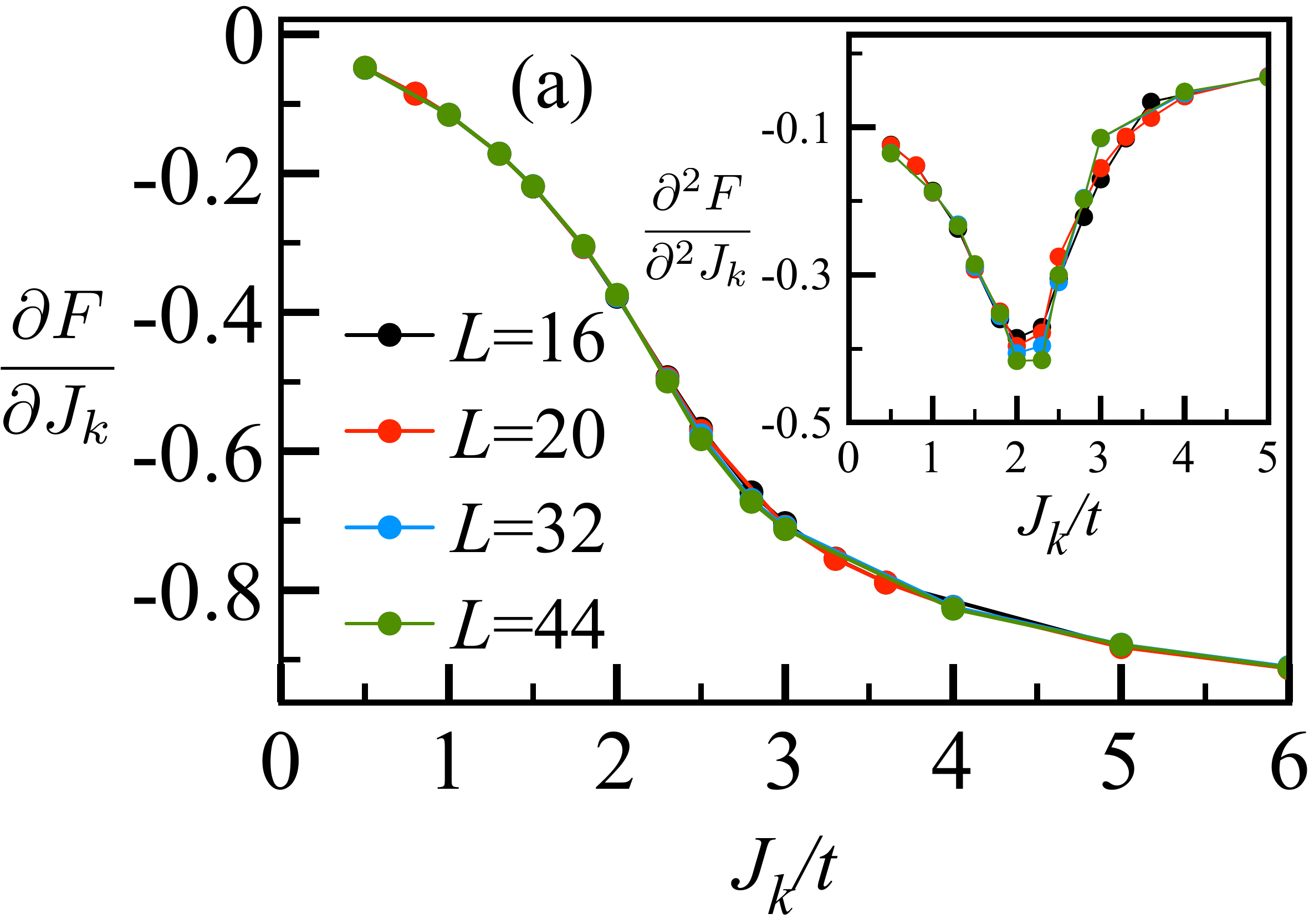}
\includegraphics[width=0.2345\textwidth]{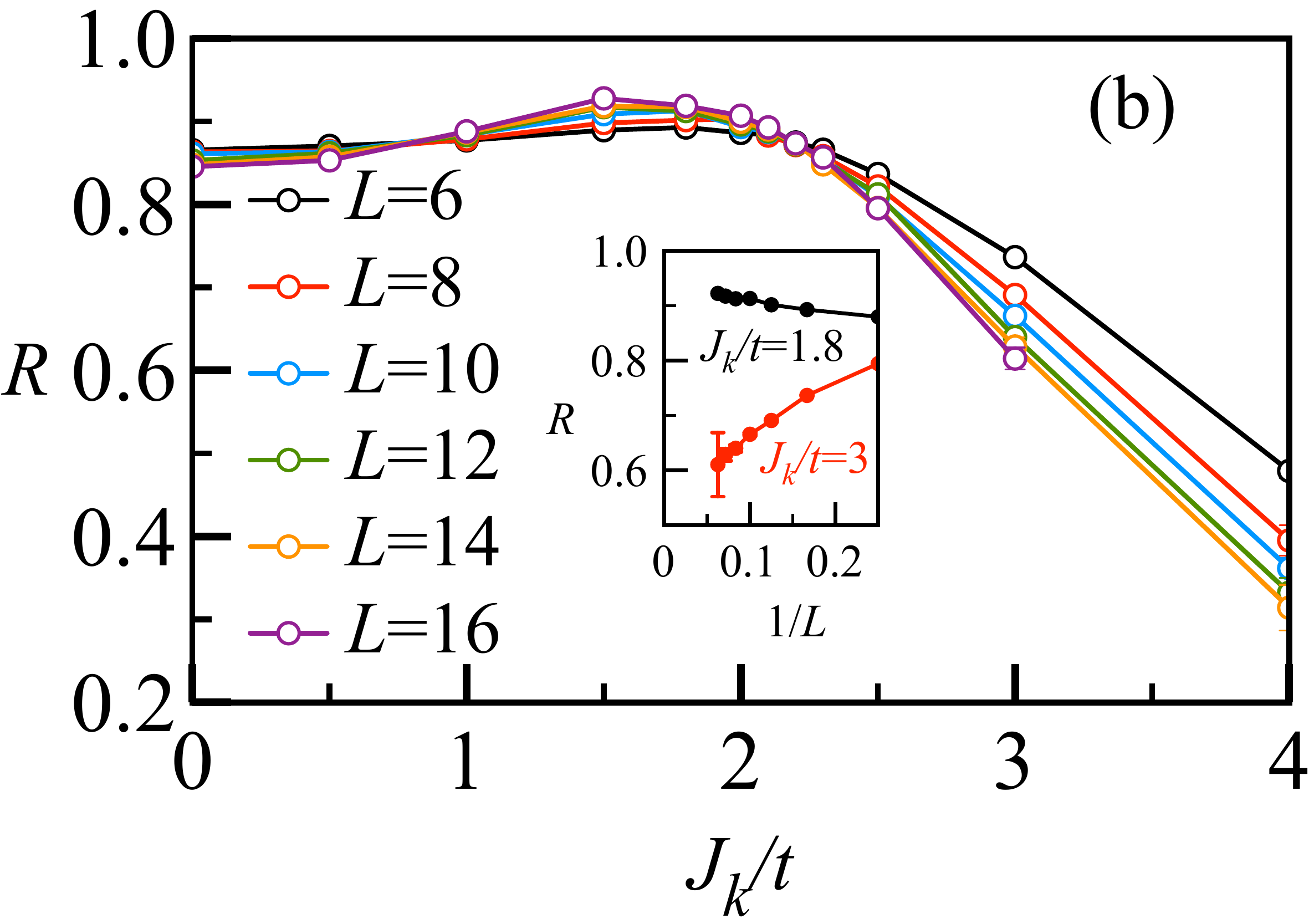}
\caption{ (a) First derivative of free energy as a function of $J_k/t$ at $\beta t=L$ and $J_h/t=1$. The inset  plots the second derivative of the free energy. (b) Correlation ratio $R$ as a function of  $J_k/t$ at $\beta t=L^2/2$ and $J_h/t=1$.  The inset   plots $R$ as a function of $1/L$ at $J_k/t=1.8$ and  $J_k/t=3$. }
\label{R_dFbydJk_vs_Jk}
\end{figure}

\textit{QMC Results. }
Given the above considerations,  we anticipate an order-disorder   transition as a function of $J_k$. To    locate it,    we  consider
  $\frac{1}{L} \frac{\partial F}{\partial J_k}=\frac{2}{3L} \sum_{\ve{r}} \langle \hat{\ve{c}}^{\dagger}_{\ve{r}}  \ve{\sigma} \hat{\ve{c}}^{}_{\ve{r}}   \cdot \hat{\ve{S}}_{\ve{r}}\rangle$ as a function of $J_k$   as  well  as  $\frac{\partial^2 F}{\partial J^2_k}$ (Fig.~\ref{R_dFbydJk_vs_Jk}(a)  and  inset).  As  apparent,
   the  data  is  consistent  with a  single  transition at $J_k^c/t \simeq  2.1$    for  $J_h/t = 1$.   Next, we  consider the
    spin susceptibility,
\begin{equation}
	\chi(\ve{k}, i \Omega_m ) =     \int_{0}^{\beta} \text{d}\tau \sum_{\ve{r}}   e^{i (\Omega_m \tau - \ve{k} \cdot \ve{r} )}   \langle   \hat{S}^z_{\ve{r}}(\tau)\hat{S}^z_{\ve{0}}(0) \rangle,
\label{chikomega.eq}
\end{equation}
from  which  we  can  define  the correlation ratio,
\begin{eqnarray}
R =1-\frac{\chi{(\ve{Q}-\delta \ve{ k} },0 )} {\chi(\ve{Q},0)} \, ,
\end{eqnarray}
where   $\ve{Q} = (\pi/a,0) $  corresponds   to  the  antiferromagnetic wave  vector  and $ \delta \ve{ k} $
 to the smallest  wave  vector on the $L$-site  chain.
  This correlation ratio   scales  to unity (zero)   for   ordered (disordered)  states,  and   at  criticality, is a  renormalization group  invariant quantity. Here,
$R= f(\left[J_k -  J_k^{c} \right] L^{1/\nu}, L^{z}/\beta, L^{-\omega} ) $ where $\nu $  is the correlation  length  exponent,  $ z$ is the  dynamical
exponent, and $\omega$  captures corrections  to scaling.      Figs.~\ref{LogCr_Stau_vs_r_tau_Jk2p}  (c) and  (d)     shows  that in the vicinity of the  critical point
 spatial  correlations   drop off  as  $1/r$    whereas    along  the imaginary time,  we  observe a  much  slower  $1/\sqrt{\tau} $    decay.
  This  suggests   a  critical exponent  $z\simeq 2$.   With this in mind,  we  can  compute  $R$ adopting a $ \beta t = L^2$  scaling  such  that
   under the assumption  vanishing correlations to scaling,   $R $  should  show a  crossing point as a function of system size  at $J_k^c$.   As
   apparent  from Fig.~\ref{R_dFbydJk_vs_Jk}(b)   $R$  shows a crossing   at  $J_k^{c}/t \simeq 2.1$  thus  providing  a consistency
   check for our  choice of the dynamical  exponent.     We  now   discuss  the  physics  at weak, $J_k  < J_k^c$,  and  strong coupling, $J_k > J_k^c$.

For $J_k  < J_k^c$, we expect-dissipation induced long-range AFM ordering. As discussed in \cite{suppl}, in this phase one can decompose the fluctuating AFM field $\ve n$ as $\ve{n}(r,\tau) = \left(\ve{\sigma}(r,\tau), \sqrt{1-\ve{\sigma}(r,\tau)^2}\right)$ where the ordering is assumed along the $\hat{z}$ direction. The low-energy action for the transverse fluctuations $\ve \sigma$ has dynamical exponent $z=2$, and is given by ${\mathcal S}_{0}(\ve \sigma) =  \frac{\Gamma}{2} \int d\tau d\tau^\prime dr\,\,\frac{{\ve{\sigma}}({r,\tau}).{\ve{\sigma}}({r,\tau'})}{(\tau - \tau')^2} +   \frac{\rho_s}{2} \int d\tau  dr\,\, (\partial_r \ve{\sigma}(r,\tau))^2$. This implies that while the $n^z$ correlations are long-ranged in both space and time, the correlations of $\ve \sigma$ are given as: $\langle \ve{\sigma}(r, \tau)  \cdot \ve{\sigma}(r, 0) \rangle \sim 1/\sqrt{\tau}$, while $\langle \ve{\sigma}(0, \tau) \cdot \ve{\sigma}(r, \tau) \rangle \sim 1/r$.

On the numerical front, at  $J_k/t = 1.8  < J_k^c /t $,   we  observe   a  slight  increase in the correlation ratio  (Fig.~\ref{R_dFbydJk_vs_Jk}(b)   inset) thus hinting to  the onset of long-range order,   but   as  $J_k$   decreases
 further  no sign of long-range order  on  our  finite lattice sizes is apparent.    To understand  this apparent lack of ordering,  we  can switch off the Kondo screening and retain only local  dissipation,  corresponding to  Eq.~(\ref{action_ef})  with  $ \chi^0 (\ve{r},  \tau)   \propto \delta_{\ve{r},\ve{0}}/\tau^2$.  For this bosonic   model  stochastic series   approaches   for  retarded interactions~\cite{Weber17,Weber21a} can   be used to  investigate this model  with  unprecedented  precision~\cite{Weber2021}.
It was shown  in Ref.\cite{Weber2021} that the marginally  relevant  nature of the  Ohmic  dissipation  at the LL  critical point    requires lattice sizes $  L  \ll L_c  \propto  e^{ \xi/J_k^2}  $   to detect long-range  order.

For  $L \lesssim L_c $   one observes  crossover  phenomena   characterized by  a  $1/r$  decay  of  the  real space spin-spin correlations and  breakdown of   Lorentz  symmetry.     Our understanding  is  that our  data  falls in this crossover  regime,  and   that for $r \ll L_c$ it  can  be  accounted for by
\begin{equation}
 \label{Space_time.eq}
 C(r,\tau)  \propto \left\{
 \begin{array}{lr}
     \frac{1}{ \sqrt{   r^2   +  \tau^{2/z} }} &     \tau \ll  \frac{1}{\Delta} \\
     \frac{e^{-\Delta  \tau}}{ \sqrt{   r^2   +  \Delta^{-2/z} }}  &   \tau \gg \frac{1}{\Delta}
 \end{array}
    \right.
\end{equation}
on an  $L$-site   lattice.  Here  $  C(r,\tau) =    e^{i\ve{Q} \cdot  \ve{r}}  \langle   \hat{S}^z_{\ve{r}}(\tau)\hat{S}^z_{\ve{0}}(0) \rangle $ and   $\Delta \propto  \left( \frac{2 \pi}{L} \right)^z $
 corresponds  to the  finite-size   gap.    At  $J_k/t = 0.5$,  \textit{far}  from the critical point, Fig.~\ref{LogCr_Stau_vs_r_tau_Jk2p}    plots   $C(r,0) $ (a) as  well  as  $ C(0,\tau) $ (b).
The  real-space  equal-time decay is  consistent with  a  $1/r$  law.  Along  the  imaginary time  we  observe crossover phenomena:  While at  short times,   $ \tau   t  \lesssim  L  $, the  temporal  decay  is consistent  with Eq.~(\ref{Space_time.eq})   at $z \simeq 1$  akin to the Heisenberg model,  we  observe   at  large  $\tau$ a  breakdown of Lorentz  invariance with $C(0,\tau)$ decaying   substantially  slower  than $1/\tau$. In the infinite-size limit,  we  foresee  that  both the  real-space  and  imaginary-time  correlations will level off to  show  long-ranged correlations,  albeit  with a very  small local moment.

 \begin{figure}[htbp]
\centering
\includegraphics[width=0.49\textwidth]{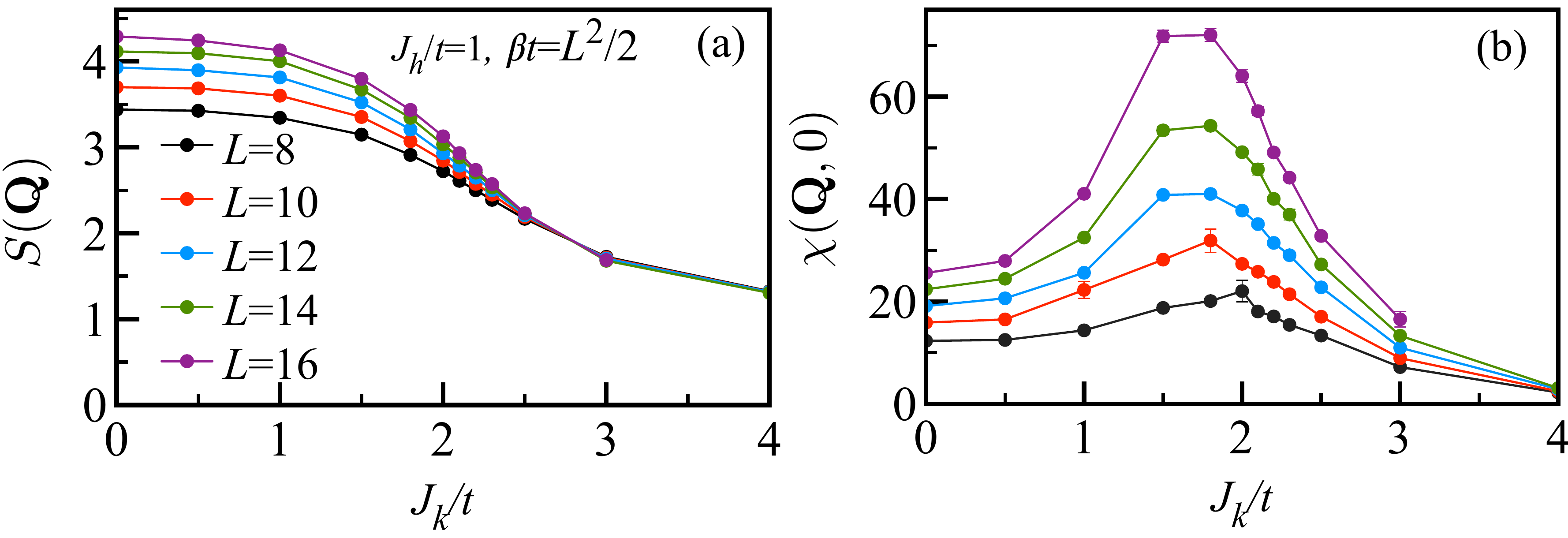}
\caption{(a) Static spin structure factor $S(\Q)$ along  the spin chain  as a function of $J_k/t$ for given $L$ at  $J_h/t=1$ and $\beta t =L^2/2$. (b) Correspondingly, spin susceptibility $\chi(\Q,0)$ as a function of $J_k/t$.}
\label{Sk_chi_eqpi_vs_Jk_L12}
\end{figure}

 The  breakdown  of Lorentz  invariance   is equally  apparent in the   data of  Fig.~\ref{Sk_chi_eqpi_vs_Jk_L12}.   The Ansatz  of  Eq.~(\ref{Space_time.eq})
 leads  a    structure  factor   $S(\ve{Q}) = \frac{1}{\beta}    \sum_{\Omega_m}   \chi(\ve{Q}, i \Omega_m)$  that   is independent on the dynamical  exponent  and  as for  the    Heisenberg  chain diverges as $\log(L)$.      The  size  scaling  in the    crossover   regime (Fig.~\ref{Sk_chi_eqpi_vs_Jk_L12}(a))     does  not show marked  differences  from the  Heisenberg  limit.  As  noted in Ref.~\cite{Weber2021}    and  seen in Fig.~\ref{Sk_chi_eqpi_vs_Jk_L12}(a),  coupling
 to the bath    reduces    the magnitude of the equal  time  spin-correlations.     On the other hand,   the   susceptibility,   $ \chi(\ve{{Q}},0)$     shows  marked  differences as a  function
 of  $J_k$.  In  Fig.~\ref{Sk_chi_eqpi_vs_Jk_L12}(b)  we  consider  the  scaling $ \beta t = L^2/2$.  In the Heisenberg  limit  this leads  to  $\chi(\ve{Q},0) \propto L$  and  a  marked  deviation from this law  is observed in the  crossover  regime.    For   $z=2$   akin to the  critical point,
 $J^c_k / t \simeq  2.1 $,    the Ansatz of  Eq.~(\ref{Space_time.eq}) yields   $ \chi(\ve{Q},0) \propto L^2 $.   This scaling  law  is  supported  by the data  thus   confirming $z \simeq 2$ at   criticality.

  \begin{figure}[htbp]
\centering
\includegraphics[width=0.49\textwidth] {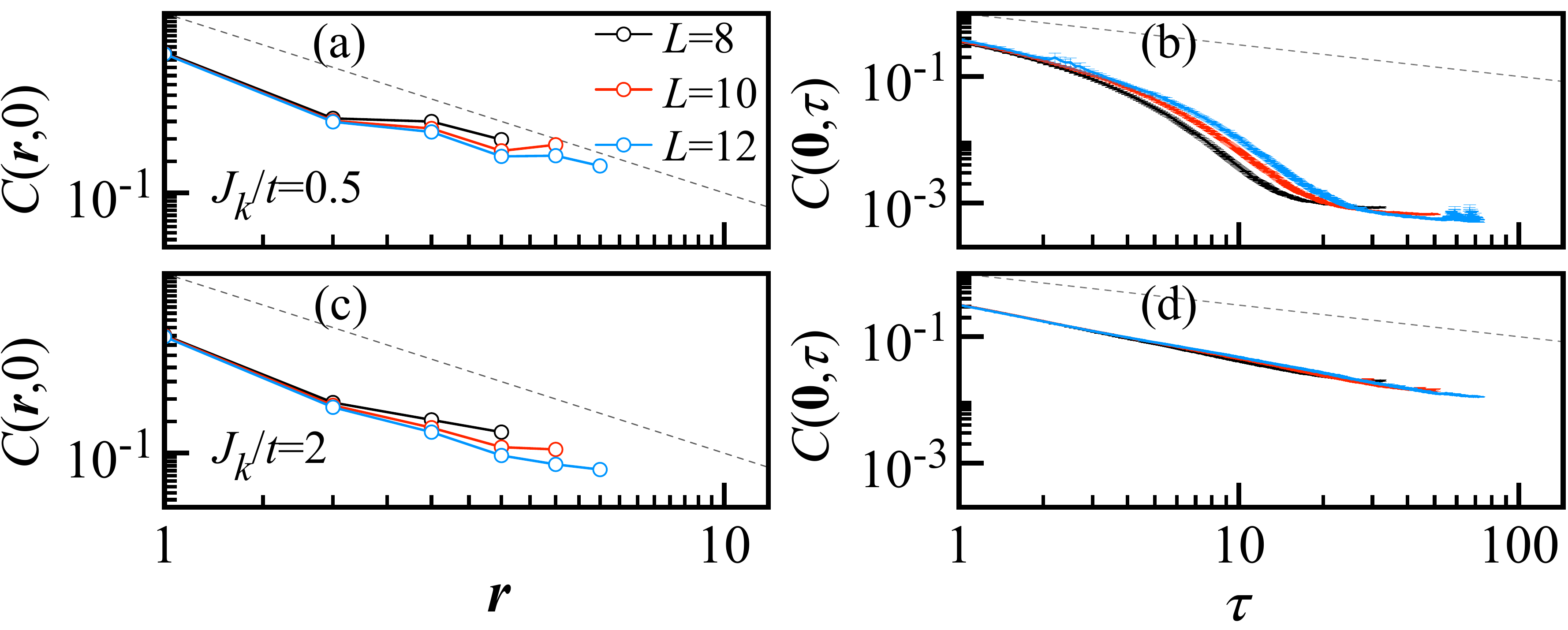}
\caption{Space and time correlation functions   $C(\ve{r},\tau) $ at  $\beta t =L^2$ and $J_h/t=1$.
(a)  $C(\ve{r},0)$ at $J_k/t=0.5$.  (b)  $C(\ve{0},\tau)$  at $J_k/t=0.5$.  (c) $C(\ve{r},0)$  at $J_k/t=0.5$.      (d)  $C(\ve{0}, \tau)$  at $J_k/t=2$.
The dashed grey lines denote the $1/r$   ((a) and (c)) and  $1/\sqrt{\tau}$  ((b) and (d)) power  laws.}%
 \label{LogCr_Stau_vs_r_tau_Jk2p}
\end{figure}

 \begin{figure}[htbp]
\centering
\includegraphics[width=0.49\textwidth]{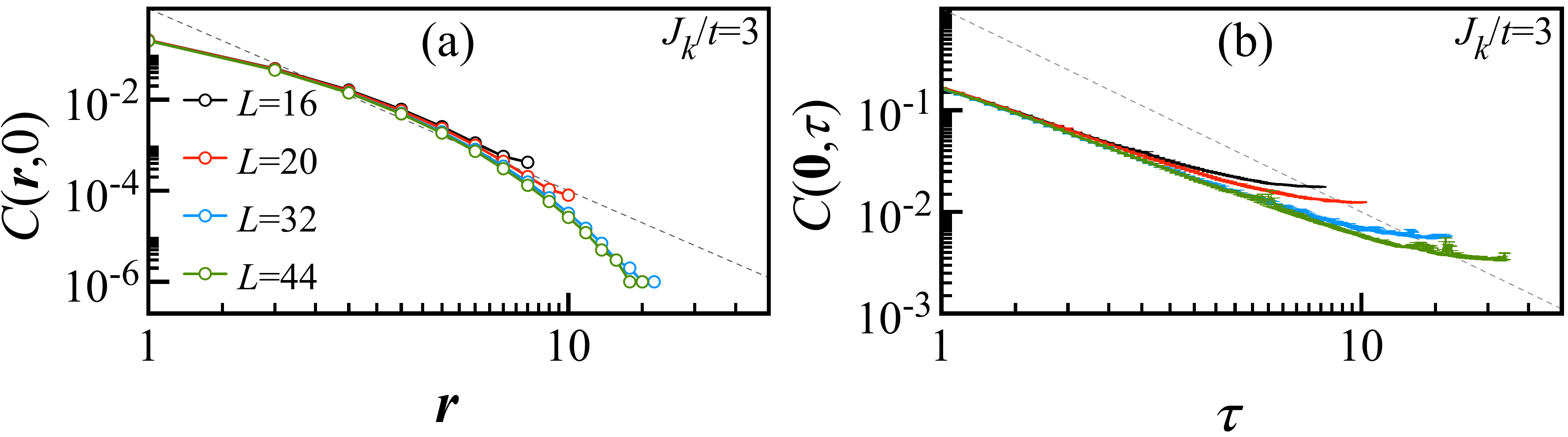}
\caption{Space and time correlation functions along the spin chain   in the Kondo-screened phase  at  $J_h/t=1, J_k/t=3$ and $\beta t =L$. (a)  $C(\ve{r},0)$.
The dashed grey line indicates  a   $1/r^4$  power   law.  (b) $C(\ve{0}, \tau)$. The dashed grey line indicates a $1/\tau^2$  law.   Both  power laws in time and in space
are  observed   in the large-$N$ limit (see Ref.~\cite{suppl}).}%
 \label{LogCr_Stau_vs_r_tau_Jk3p}
\end{figure}

At $J_k  >J_k^c$   we  are in a Kondo  screened  phase   that can be understood  within a   large-$N$  mean field  theory  presented in Ref.~\cite{suppl}. In this phase, the  spin-spin correlations  inherit  the    asymptotic  behavior  of  the  conduction electrons  and   fall off  as $1/r^4$ in  space  and as $1/\tau^2$  in  imaginary  time.  In particular, Figs.~S2 and S3 of Ref.~\onlinecite{suppl} plot  the space- and time-displaced correlation functions  within the large-$N$  approximation and  confirm  the above.  The QMC  data of Fig.~\ref{LogCr_Stau_vs_r_tau_Jk3p}   is consistent  with this expectation.

 \begin{figure}[htbp]
\centering
\includegraphics[width=0.46\textwidth] {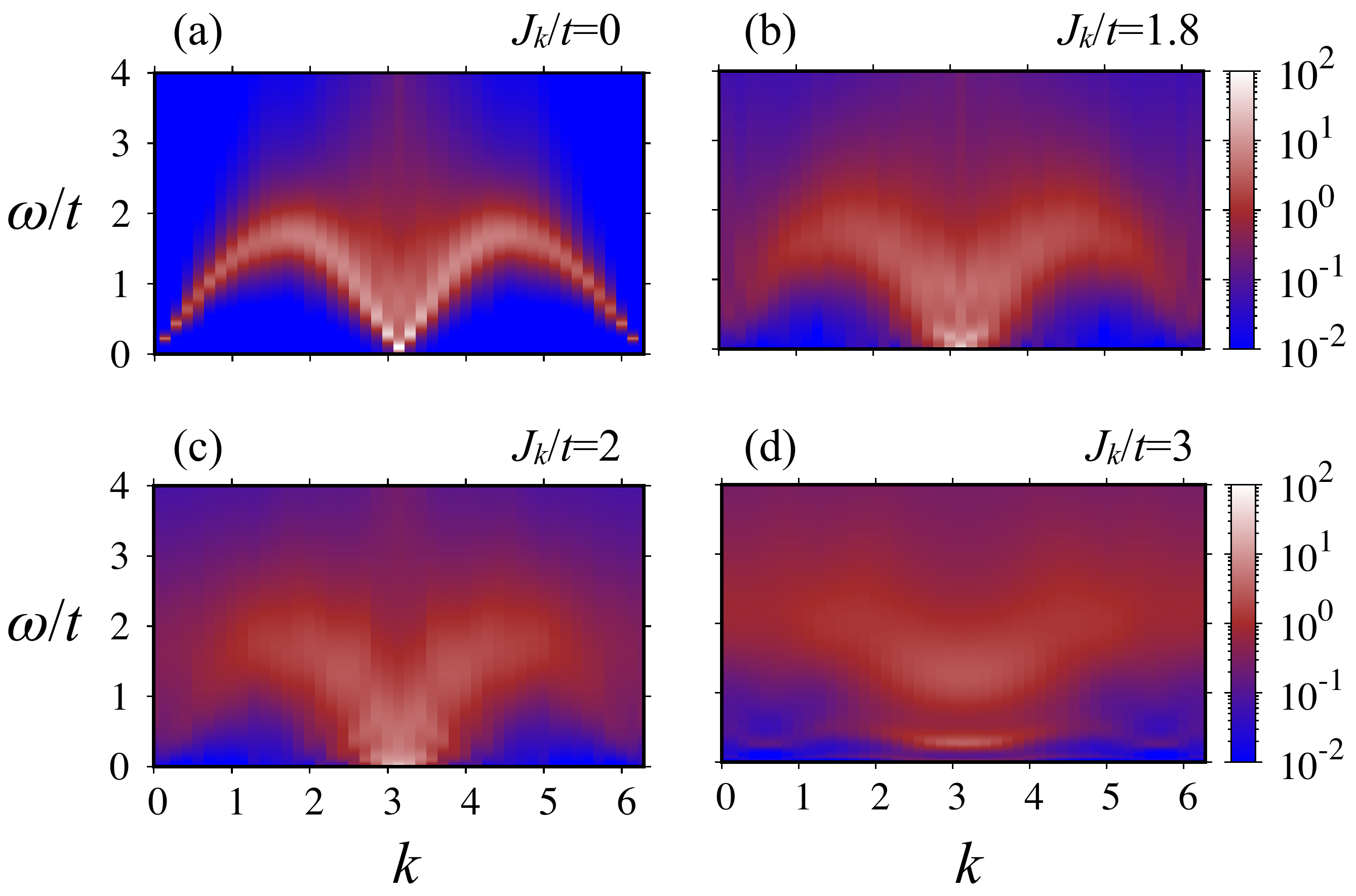}%
\caption{$S(k,\omega)$ as a function of energy $\omega/t$ and momentum $k$ along the spin chain at $\beta t=L=44$ and  $J_h/t=1$.}
\label{Skomegavskomega_L44}
\end{figure}

Using  the ALF~\cite{ALF_v2} implementation of the maximum Entropy  method~\cite{Sandvik98, KBeach2004} we  compute  the
dynamical  spin  structure  factor $S(\ve{k},\omega)=\frac{\text{Im} \chi (\ve{k}, \omega + i0^+) }{1-e^{-\beta \omega}}$.
 Fig.~\ref{Skomegavskomega_L44}(a)  plots  this
quantity  for   the Heisenberg model.  The data  shows the  well known  two   spinon continuum~\cite{Cloizeaux62, Mueller81,Caux05}.
 At finite   Kondo couplings (Figs.~\ref{Skomegavskomega_L44}(b)-(c)), the  two-spinon continuum   is  still apparent at elevated energies.
 However  the  low-energy  bound  shows a   marked  deviation  from the  linear   dispersion  and is  very suggestive of a $\omega\propto k^2$  law. In fact, a field  theory presented in Ref.~\onlinecite{suppl} as  well as  a  large-$S$ calculation~\cite{Weber2021} of a  Heisenberg   chain locally    coupled to an Ohmic bath  confirms that dissipation  stabilizes long-range  order   and  that   the  lower bound  of  the dispersion relation  follows  an  $\omega\propto k^2$  law  akin to Landau-damped  Goldstone modes.
 Our  dynamical  data    bears  similarities with   spinon  binding  as observed in KCuF$_3$~\cite{BellaLake2005}  and corresponding to a  dimensional crossover~\cite{Raczkowski13}.  In the present case, the elevated-energy spectrum  shows the   two-spinon  continuum   while  the  low energy to corresponds the spin-wave   excitations of the   Heisenberg  chain coupled to an Ohmic bath~\cite{suppl}. Finally, in the Kondo-screened phase at $J_k/t=3$ see Fig.~\ref{Skomegavskomega_L44} (d)  the low-lying spectral weight  is depleted.

 \begin{figure}[htbp]
\centering
\includegraphics[width=0.49\textwidth] {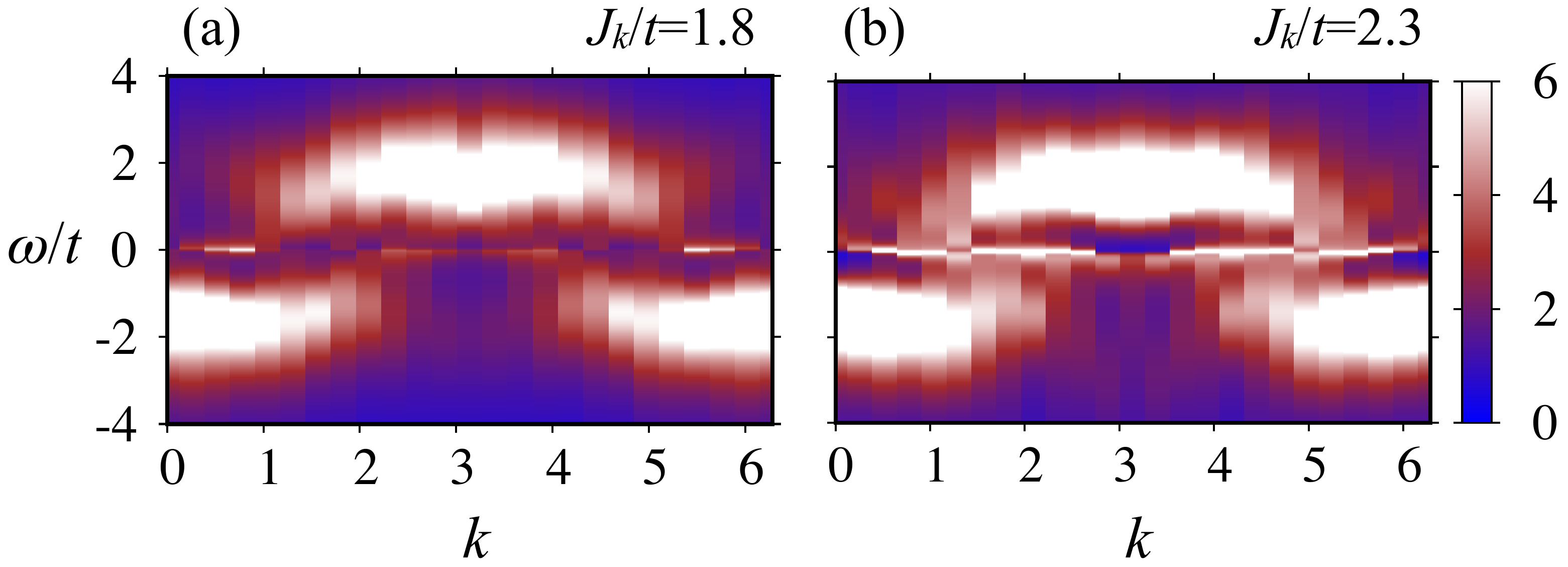}  \\

~

\includegraphics[width=0.46\textwidth]{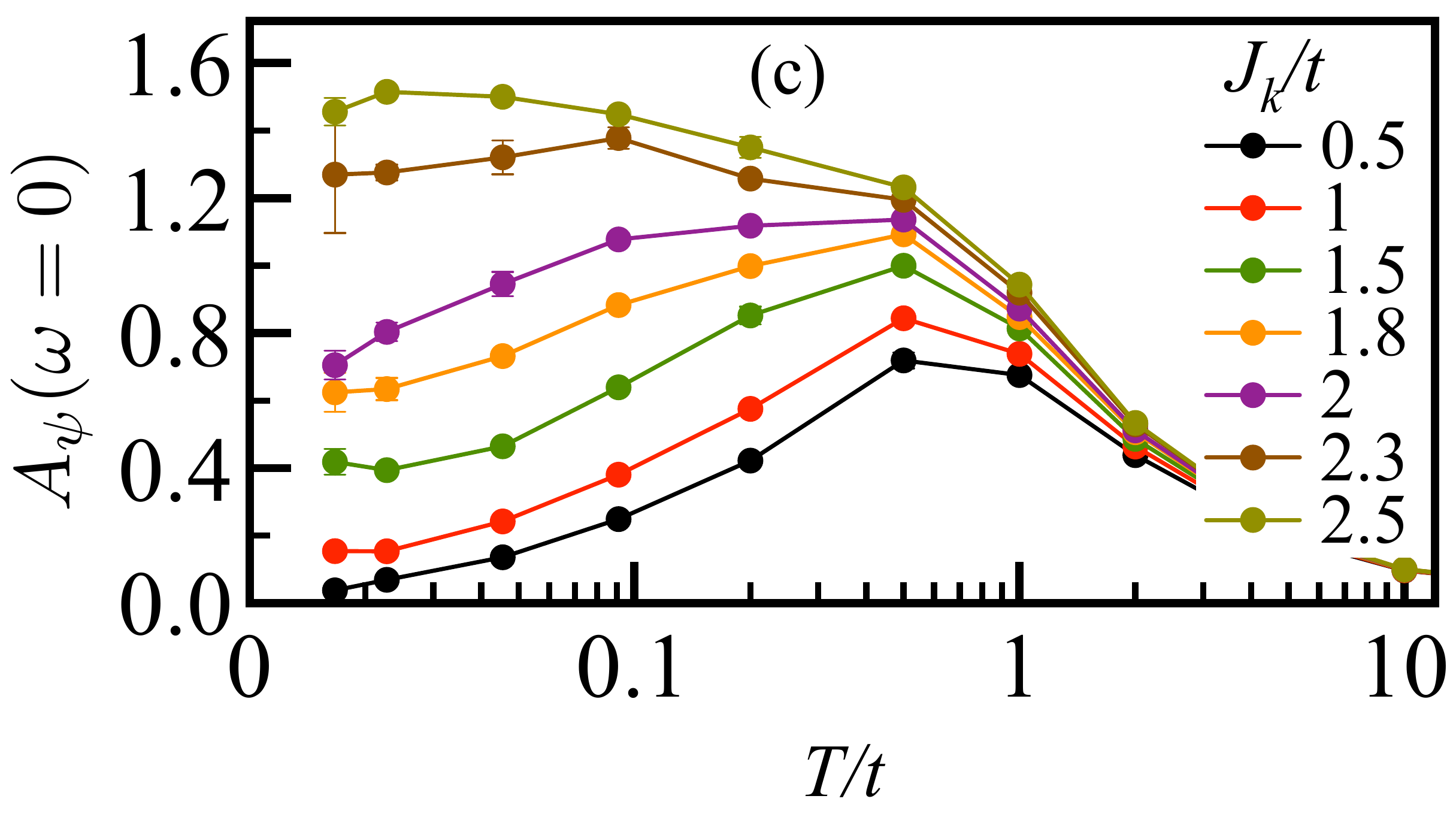}
\caption{
Spectral function of the composite-fermion operator $A_{\psi}(k,\omega)$ on an $L=24$ lattice at  $\beta t = 48$ (a) in the ordered  phase   and  (b) in the Kondo-screened  phase.
(c) Local zero-bias signal  $A_\psi(\omega=0)$ as a function of temperature $T/t$  at $L=44$ and for various values of $J_k/t$ in the ordered  and disordered phases.
}
\label{Apsikom.fig}
\end{figure}

We  now turn our  attention to  Kondo  screening    and  heavy-fermion physics.    Consider  the composite-fermion  operator  $\hat{\psi}^\dagger_{\ve{r},\sigma}=2\sum_{\sigma^\prime} \hat{c}^\dagger_{\ve{r},\sigma^\prime} \ve{\sigma}_{\sigma^\prime, \sigma} \cdot \hat{\ve{S}}_{\ve{r}}$~\cite{Costi00,Borda07,Raczkowski18}.  In the  large-$N$ limit   this   quantity  picks  up the Higgs  condensate   or  hybridization matrix  element~\cite{suppl},   characteristic   of Kondo screening~\cite{Danu2021}. Here we compute the spectral function $A_{\psi}(\ve{k},\omega) =- \text{Im}  G^{\text{ret}}_{\psi} (\ve{k},\omega )$ with  $G_{\psi}^{\text{ret}}  (\ve{k}, \omega) = - i \int_{0}^{\infty} dt e^{i \omega t}  \sum_{\sigma } \langle \big\{\hat{\psi}_{\ve{k},\sigma}^{}(t), \hat{\psi}_{\ve{k},\sigma}^{\dagger} (0) \big\} \rangle$, representing the conduction-electron T matrix.
Figs.~\ref{Apsikom.fig}(a)-(b)   shows  flat  (i.e  heavy)  bands in the vicinity of the Fermi energy,   both  below  and  above the critical point $J_k^c/t\simeq 2.1$. In  Fig.~\ref{Apsikom.fig}(c)   we  plot  $A_{\psi}(\omega)   =  \frac{1}{L} \sum_{\ve{k}} A_{\psi}(\ve{k}, \omega) $  as a function of  temperature and  $J_k$.   To avoid  analytical  continuation,  we   use the relation $ A_{\psi}(\omega=0)   \simeq(1/\pi) \beta G_{\psi}(\tau=\beta/2) $. Confirming the  $\ve{k}$-dependent data,  we   see that this quantity never  vanishes  at  low temperatures in  both  phases.  Hence, the  data  supports  the point of  view  that Kondo    screening  is  active in the  dissipation-induced  ordered  phase.
It  is  interesting to note  that  the  temperature  dependence of    $A_{\psi}(\omega=0)$   differs in  both phases.    While it  grows  and  saturates in the Kondo-screened phase,     it  shows a maximum  in the ordered phase.  Such a behavior  can  be understood in terms of the onset of ordering that  opens a pseudogap in the  spectral function, see Ref.~\onlinecite{suppl}.
$A_{\psi}(\omega=0)$   is an important  quantity  since it  provides a  link to  STM  experiments.  In   fact,  it  corresponds  to the  zero bias signal $dI_{\ve{l}}/dV(V=0)$  for   tunneling  processes between the  tip  and the  substrate  that  involve intermediate  excited  states of  the  localized orbital. In  the  experiments  described in Refs.~\cite{Spinelli2015, Toskovic2016}  and  modelled in  Ref.~\cite{Danu2019}  $J_k$ can be tuned by changing the width of  the Cu$_2$N islands between the Co adatoms and Cu(100) surface.   Provided  that the  chains are  long  enough, our  observation of   distinct  temperature   behaviors  of $A_{\psi}(\omega=0)$ in the  two  phases  provides  a  means to  experimentally distinguish them.

\textit{Conclusions.}
The  physics of  the  Heisenberg  chain  coupled to   two-dimensional  electrons   can be  understood  by the competition between  dissipation and  Kondo screening. Not unlike the competition between the  RKKY  interaction and Kondo screening, the  Kondo  coupling triggers  both effects but at different energy scales.

At  weak coupling  dissipation  dominates,  and the   chain  develops long-range  antiferromagnetic order. Since  the  coupling to the Ohmic bath is marginally  relevant~\cite{Weber2021},  system sizes  exceeding  our achievable  lattices  are  required to unambiguously detect the order.   As   a  consequence our data  below the  critical point are  dominated by a crossover  regime  characterized by the absence of Lorentz invariance as seen in Ref.~\onlinecite{Weber2021}.  In fact, the tendency towards ordering on small  lattices can be enhanced by  considering the XXZ (as opposed to Heisenberg) chain in its Luttinger-liquid  phase. Here  the  transverse spin  operator   acquires   a scaling  dimension  smaller than  1/2 such that the coupling to the Ohmic bath becomes relevant.  In Ref.~\onlinecite{suppl} we  have  verified that the  XXZ  chain indeed  leads to  a  stronger  tendency  towards  ordering. At  strong  coupling,  Kondo screening dominates, leading to  a  paramagnetic  phase.  Here, the  spin-spin correlations  inherit the power-law  decay  of  the   host metal.   Aspects of  this  phase  diagram  have  been put  forward  based on analytical  considerations in  Ref.~\onlinecite{lobos2012magnetic}.

The composite-fermion spectral function  reveals a heavy  band in both phases.  Hence,  in the ordered phase  Kondo screening coexists with  dissipation-induced  ordering. In the disordered  phase,  the  notion of  heavy Fermi liquid  can be made  precise by invoking the Luttinger theorem that relates the size of the Fermi surface to the count of its constituents. Since  our model preserves translation invariance along the chain,  it can be thought of as a one-dimensional model  with a unit cell that contains $L$  conduction electrons and single spin-1/2 local moment. With this formulation, one can now apply  Oshikawa's   proof of  Luttinger  theorem
in the context of Kondo lattices~\cite{Oshikawa00a},  and show that the volume of the Fermi surface includes the local moments, assuming that the system is described by a Fermi liquid at low energies.  An explicit calculation is  presented in Ref.~\onlinecite{suppl}.

Our model provides  a  unique negative-sign-free realization of a Hertz-Millis-type  transition between  antiferromagnetic and paramagnetic  heavy-fermion  metals.  To  avoid  the  negative-sign problem  we  have to  consider  a particle-hole  symmetric  conduction band   with  inherent  nesting  instabilities.  Nevertheless, we expect that the our broad conclusions, including the global structure of the phase diagram, hold  for  a  generic  two-dimensional Fermi  surface.  In  the latter  case,  the  spin-spin correlations of  the host metal  decay   as  $1/r^3$   and as  $1/\tau^2$ in space and time, respectively.  Since the $1/r^3$ decay is an irrelevant perturbation at  the   decoupled fixed  point,   we  expect   dissipation-induced  ordering   to occur generically at  small $J_k$.
Our numerical data suggests that the transition in Fig.~\ref{phsgm_RG_QMC} is characterized  by  a  dynamical exponent $z \simeq 2$.  A detailed understanding  of the  transition,  both on the  numerical and analytical fronts, remains for  future work.

\acknowledgments

FFA  thanks  D. Luitz and  M. Weber  for   discussions   on  the   dissipative  Heisenberg  chain.
The authors gratefully acknowledge the Gauss Centre for Supercomputing e.V. (www.gauss-centre.eu) for providing computing time on the GCS Supercomputer SUPERMUC-NG at Leibniz Supercomputing Centre (www.lrz.de). The research has been supported by the Deutsche Forschungsgemeinschaft through grant number AS 120/14-1 (FFA), the W\"urzburg-Dresden Cluster of Excellence on Complexity and Topology in Quantum Matter - ct.qmat (EXC 2147, project-id 390858490) (FFA and MV), and SFB 1143 (project-id 247310070) (MV). TG is supported by the National Science Foundation under Grant No. DMR-1752417, and as an Alfred P. Sloan Research Fellow. FFA and TG thank the BaCaTeC for partial financial support.

\bibliography{Kondo_ref1,Kondo_ref2,fassaad}
\clearpage
\newpage
\widetext
\renewcommand\theequation{S\arabic{equation}}
\renewcommand\thefigure{S\arabic{figure}}
\setcounter{equation}{0}
\setcounter{figure}{0}
\begin{center}
 \Large{ \bf Supplemental Material for:
 Spin chain on a metallic surface: Dissipation-induced order vs. Kondo entanglement }\label{supplement}

{ \small Bimla Danu,  Matthias Vojta, Tarun Grover, and  Fakher F. Assaad}
\end{center}
\section{Logarithmic correction to Hertz-Millis action due to nested Fermi surface}

The spin-susceptibility $\chi^0(\mathbf{r},\tau)$ of non-interacting conduction electrons is proportional to
$\langle \ve{c}^{\dagger}_{\ve{r}}  (\tau)\ve{\sigma} \ve{c}_{\ve{r}} (\tau)   \cdot  \ve{c}^{\dagger}_{\ve{0}}(0) \ve{\sigma} \ve{c}_{\ve{r}} (0)\rangle$. As discussed in the main text, the important contribution to the Hertz-Millis action comes only from  $\chi^0(\ve{r}=0,\tau)$. A simple calculation using Wick's theorem shows that  $\chi^0(\ve{r}=0,\tau) \sim I^2$ where
\be
I = \int_{k_x,k_y \in \textrm{F.S.}} dk_x \,dk_y\,\, e^{-\tau(\cos k_x + \cos k_y)}
\ee
where $k_x,k_y \in \textrm{F.S.}$ means that $k_x,k_y$ are inside the Fermi surface of the square-lattice nearest-neighbor tight-binding model at half-filling, i.e., the momenta $(k_x,k_y)$ are occupied at $T = 0$. Since the Fermi surface consists of flat portions at angles $\pm \pi/4$, it is convenient to consider the integral $I$ in a rotated coordinate system, $k'_x = (k_x + k_y)/\sqrt{2}, k'_y = (-k_x + k_y)/\sqrt{2}$. Utilizing the symmetry of the integrand and rescaling the variables, the integral now becomes

\begin{equation}
	I =  8 \int_{0}^{\pi/2} dk'_x \, \int_{0}^{\pi/2} dk'_y\,\, e^{-\tau \cos k'_x \cos k'_y}
\end{equation}

Since we are interested in the limit $\tau \gg 1$, we can evaluate the above integral using  the  saddle point. There are two distinct saddle points: the first arises when $k'_x = \pi/2$ and $k'_y$ is not close to $\pi/2$ -- these are the boundaries of the Fermi surface that \textit{exclude} the four corner points $(k_x, k_y) = (\pm \pi,0), (0, \pm \pi)$ of the diamond in original $(k_x, k_y)$ coordinate system. The second set of saddles precisely correspond to these four corner points. These two set of saddle points yield qualitatively different contribution to $I$, which we denote by $I_1$ and $I_2$ respectively.

Let us first consider contribution to from saddles along $(k'_x, k'_y) = (\pi/2, k'_y)$ where $k'_y$ is not close to $\pi/2$. There will be identical contribution from saddles  $(k'_x, k'_y) = (k'_x, \pi/2)$ with $k'_x$ not close to $\pi/2$. Close to these saddles, we can write $k'_x = \pi/2 - x$, so that the contribution $I_1$ to the integral is approximately given by

\begin{equation}
	I_1 \sim  \int_{0}^{c_1} dx \, \int_{0}^{\pi/2-c_2} dk'_y\,\, e^{-\tau \sin x \cos k'_y}
\end{equation}
where $c_1 \ll 1$ is a small number, and $c_2 = O(1)$. The precise values of $c_1, c_2$ are irrelevant for asymptotic analysis. Since $c_1 \ll 1$, we can expand $\sin(x) \sim x$, and perform the integral over $x$ leading to $I_1 \sim  \int_{0}^{\pi/2-c_2} dk'_y\,\, \frac{1}{\tau \cos k'_y}$. The integral over $k'_y$ converges and yields an $O(1)$ prefactor because by construction, $k'_y$ is never close to $\pi/2$. Therefore, $I_1 \sim 1/\tau$.

Next, consider $I_2$, i.e., contribution to $I$ from  $(k'_x, k'_y) \approx (\pi/2, \pi/2)$. Writing $k'_x = \pi/2 - x, k'_y = \pi/2 - y$, where $x, y \ll 1$, one finds

\begin{equation}
	I_2 \sim \int_{0}^{a} dx \, \int_{0}^{a} dy\,\, e^{-\tau x y} \sim \int_{0}^{a} dy \left( \frac{1 - e^{-a \tau y}}{\tau y}\right)
\end{equation}
where $a \lesssim 1$ is some constant. Since $\tau \gg 1$, the integrand for the integral over $y$ has two regimes, the first corresponding to $y \lesssim 1/(\tau a)$ and the other $y \gtrsim 1/(\tau a)$. Therefore, we split the integral as

\begin{equation}
	I_2    \sim \int_{0}^{1/(\tau a)} dy \left( \frac{1 - e^{-a \tau y}}{\tau y}\right) + \int_{1/(\tau a)}^{a} dy \left( \frac{1 - e^{-a \tau y}}{\tau y}\right) \equiv I_{2,A} + I_{2,B}
\end{equation}

$I_{2,A}$ is given by

\begin{equation}
	I_{2,A} \sim  \int_{0}^{1/(\tau a)} dy \left( \frac{1 - e^{-a \tau y}}{\tau y}\right) \sim 1/\tau
\end{equation}
where we have obtained the asymptotic dependence by simply Taylor expanding the integrand, which is allowed because in the range of the integral, $y \ll 1$ (since $1/(\tau a) \ll 1$). In contrast, the leading contribution to $I_{2,B}$ is

\begin{equation}
	I_{2,B} \sim     \int_{1/(\tau a)}^{a} dy \left( \frac{1 - e^{-a \tau y}}{\tau y}\right) \sim \frac{1}{\tau} \log(a^2 \tau) \sim \frac{\log(\tau)}{\tau}
\end{equation}

Combining everything, at the leading order, $I \sim I_1 + I_2 \sim  \frac{\log(\tau)}{\tau}$. Therefore, the leading term in the Hertz-Millis action is proportional to $J^2_k \int d\tau d\tau^\prime \sum_{\ve{r}}  \frac{\log^2(\tau - \tau')}{\left(\tau-\tau'\right)^2}  {\ve{n}}_{\ve{r}} (\tau) \cdot {\ve{n}}_{\ve{r}}(\tau^\prime) $, i.e., compared to the case of non-nested Fermi surface, one obtains a multiplicative logarithmic correction.

\section{Spin autocorrelations for a nested Fermi surface}

For a generic, non-nested Fermi surface in two spatial dimensions,  the AFM correlations decay as: $\langle \ve{n}(\ve{r}) \cdot \ve{n}(\ve{0}) \rangle \sim 1/r^3$.  However, as discussed in the main text, in the paramagnetic phase of our model, we find that the AFM correlations both for the spin-chain and for the conduction electrons decay as: $\langle \ve{n}(\ve{r})  \cdot \ve{n}(0) \rangle \sim 1/r^4$ when $\ve{r} =  r (1,0) $ lies along the direction of the chain.  The change in the power-law compared to a generic Fermi surface is a consequence of the nested Fermi surface, and we provide a brief derivation here.

Using Wick's theorem, $\langle \ve{n}(\ve{r}) \cdot \ve{n}(\ve{0}) \rangle \sim \langle c^{\dagger}_{\ve{r},\sigma} c_{\ve{r}, \sigma} \rangle^2 \sim \left( \frac{1}{V} \sum_{\ve{k}} n_F(k_x, k_y) e^{i \ve{k}.\ve{r}}\right)^2$ where $V = L^2$ is the total system size and $n_F$ is the Fermi function at $T = 0$. Orienting the axes so that the filled Fermi sea corresponds to $k_x \in (-\pi/\sqrt{2},-\pi/\sqrt{2}), k_y \in (-\pi/\sqrt{2},-\pi/\sqrt{2})$, one may now exploit the square shape of the corresponding Fermi surface to decompose $n_F$ as $n_F(k_x, k_y) = f(k_x) f(k_y)$ where $f(k) = 1$ for $k \in   (-\pi/\sqrt{2},-\pi/\sqrt{2})$ and $f(k) = 0$ otherwise. With this coordinate choice, $\ve{r} = (\ve{x} + \ve{y})/\sqrt{2}$. Therefore, one finds
\begin{equation}
\langle \ve{n}(\ve{r}) \cdot \ve{n}(\ve{0})\rangle  \sim   \left(g(r) \right)^4
\end{equation}
where $g(r) =  \sum_{k} f(k) e^{i k r/\sqrt{2}}/L$. One now recognizes that $g(r)$ precisely corresponds to the single-particle Green's function for a 1d system with Fermi points at $\pm \pi/\sqrt{2}$, and therefore, $g(r) \sim 1/r$. Thus, one obtains the result seen in our numerics, namely,  $\langle \ve{n}(\ve{r}) \cdot \ve{n}(\ve{0}) \rangle \sim 1/r^4$. In passing, we note that if  $\ve{r}$ instead makes a $\pi/4$ angle to the spin-chain, then following the same steps, one finds that $\langle \ve{n}(\ve{r}) \cdot \ve{n}(\ve{0}) \rangle \sim 1/r^2$.

\section {Stability of ordering for a dissipative spin chain}\label{spinwaveanal}

To gain  insight into the stability of magnetic excitations on the spin chain in the presence of dissipative Kondo coupling, let us first consider a one dimensional system with the following imaginary-time non-linear sigma action for an $O(3)$ order parameter $\ve{n}$:

\begin{equation}
{\mathcal S}_{\text{diss} }(\ve n)  =  \frac{\Gamma}{2} \int d\tau d\tau^\prime dr\,\, \frac{{\ve{n}}({r,\tau}).{\ve{n}}({r,\tau'})}{(\tau - \tau')^2} +  \frac{\rho_s}{2} \int d\tau  dr\,\, (\partial_\tau \ve{n}(r,\tau))^2 + (\partial_r \ve{n}(r,\tau))^2, \label{eq:O3diss}
\end{equation}
where $\ve{n}(r,\tau) \cdot \ve{n}(r,\tau) = 1$ and $\Gamma$ is the strength of the dissipation. Let us assume that the $O(3)$ symmetry is spontaneously broken down to $O(2)$ with the order-parameter pointing along the $\hat{z}$ direction. One may then parameterize $\ve{n}$ as: $\ve{n}(r,\tau) = \left(\ve{\sigma}(r,\tau), \sqrt{1-\ve{\sigma}(r,\tau)^2}\right)$ where $\ve \sigma$ is a two-component vector that captures the transverse fluctuations of $\ve n$. Assuming $|\ve{\sigma}(r,\tau)| \ll 1$, one may then obtain the effective action ${\mathcal S}_{\text{diss} }(\ve \sigma) $ in the putative symmetry-broken phase. One finds ${\mathcal S}_{\text{diss} }(\ve \sigma) = {\mathcal S}_{0}(\ve \sigma) + {\mathcal S}_{1}(\ve \sigma) + ...$ where

\begin{equation}
{\mathcal S}_{0}(\ve \sigma) =  \frac{\Gamma}{2} \int d\tau d\tau^\prime dr\,\, \frac{{\ve{\sigma}}({r,\tau}).{\ve{\sigma}}({r,\tau'})}{(\tau - \tau')^2} +   \frac{\rho_s}{2} \int d\tau  dr\,\, (\partial_r \ve{\sigma}(r,\tau))^2,
 \end{equation}

\begin{equation}
{\mathcal S}_{1}(\ve \sigma) =  \frac{\Gamma}{8} \int d\tau d\tau^\prime dr\,\, \frac{{\ve{\sigma}^2}({r,\tau}).{\ve{\sigma}^2}({r,\tau'})}{(\tau - \tau')^2} +  \frac{\rho_s}{2} \int d\tau  dr\,\,  \left(\ve{\sigma}(r,\tau) \cdot \partial_r \ve{\sigma}(r,\tau)\right)^2,
\end{equation}
 and `...' denotes higher order terms that are less relevant than ${\mathcal S}_{0}(\ve \sigma)$ and ${\mathcal S}_{1}(\ve \sigma)$. The action ${\mathcal S}_{0}(\ve \sigma)$ is invariant under the following scaling transformation: $r \rightarrow \lambda r, \tau \rightarrow \lambda^2 \tau, \sigma(r,\tau) \rightarrow \sigma(r,\tau)/\sqrt{\lambda} $. On the other hand, under the same scaling transformation, ${\mathcal S}_{1}(\ve \sigma) \rightarrow {\mathcal S}_{1}(\ve \sigma)/\lambda$, and therefore, $\mathcal{S}_1(\ve \sigma)$ is irrelevant at the RG fixed point governed by ${\mathcal S}_{0}(\ve \sigma)$. Terms such as   $\int d\tau  dr\,\, (\partial_\tau \ve{\sigma}(r,\tau))^2 $ and  $\int d\tau  dr\,\,  \left(\ve{\sigma}.\partial_\tau \ve{\sigma}\right)^2$ are even more irrelevant at this fixed point. Therefore, the low-energy theory in the symmetry-broken phase is given by $\mathcal{S}_0(\ve \sigma)$, which has dynamical critical exponent $z = 2$ and corresponds to a Landau-damped Goldstone boson. Irrelevancy of interactions between damped Goldstone modes, such as the term $S_1(\ve \sigma)$, indicates that the ordered phase is stable against fluctuations. In contrast, in the absence of the dissipative term in Eq.~(\ref{eq:O3diss}), i.e. when $\Gamma = 0$, the interactions between the Goldstone modes would be marginal at the leading order, and lead to logarithmic divergence in various quantities, in accordance with Mermin-Wagner theorem. Dissipation alters the scaling dimension of the Goldstone mode, $\ve \sigma$, and obviates Mermin-Wagner theorem.

Another related approach to confirm the stability of the ordered phase is via calculating the reduction in the order parameter due to fluctuations. The magnitude $m$ of the order parameter is $m = \sqrt{1-\ve{\sigma}(r,\tau)^2} \approx 1 - \ve{\sigma}(r,\tau)^2/2$. Therefore, at the leading order, the reduction in the order parameter approximately equals $\langle \ve{\sigma}(r,\tau)^2\rangle/2 \sim \int dk\,d\Omega \frac{1}{\Gamma |\Omega| + \rho_s k^2}$. As one may readily verify, this integral converges in the infrared, and therefore, the fluctuations do not destroy the ordering (unlike the case of a non-dissipative chain where the analogous correction to the order parameter diverges logarithmically).

We note that the same conclusion as above can also be reached by a semiclassical, large-$S$ calculation using the Holstein-Primakoff representation for spins. Details can be found in the  supplemental material of  Ref.~\onlinecite{Weber2021}.

\section {Mean-field theory for a dissipative spin chain Kondo coupled to 2D electrons}
\label{large-mlfd}

To formulate the  mean field we consider a spin-1/2 chain  along   the $x$-axis of a square conducting substrate and impose   periodic boundary conditions both along the spin chain and on the substrate.  In this case the unit cell $\r$ contains the $n=1 \cdots  L$ conduction electrons $\hat{c}_{{\ve  r},n,\sigma }$   and a  single spin-1/2 degree of freedom $\hat{\ve{S}}_{\ve r}$. We use  the fermionic representation of local moment $\hat{\ve{S}}_{\ve r}=\frac{1}{2}\sum_{\sigma,\sigma^\prime} \hat{d}^\dagger_{{\ve r},\sigma}  \ve{\sigma}_{\sigma,\sigma^\prime} \hat{d}_{{\ve r},\sigma^\prime}$ and impose  the constraint $\sum_\sigma \hat{d}^\dagger_{{\ve r},\sigma}  \hat{d}_{{\ve r},\sigma}  =1$.

In this fermionic representation  the partition function $Z$ for the Hamiltonian given in Eq.~(1)  of main paper can be written as,
 \begin{eqnarray}\label{Zact}
 Z=\mbox{Tr} [e^{-\beta \hat{H}}] \equiv \int \mathcal{D} (\bar{\phi}, \phi, \lambda) e^{-{\mathcal S(\bar{\phi}, \phi ,\lambda) }}
 \end{eqnarray}
with  ${\phi}=(c, d \big) $  and  the action ${\mathcal S(\bar{\phi}, \phi, \lambda)}$,
 \begin{eqnarray}\label{Sact}
{\mathcal S}(\bar{\phi}, \phi , \lambda) &=&\int^\beta_0 d \tau  \sum_{{\ve r},n,\sigma}\bar{c}_{{\ve r}, n,\sigma, \tau}\partial_\tau  {c}_{{\ve r},n,\sigma,\tau} -t \int^\beta_0 d \tau\sum_{{\ve r}, n,\sigma} \big(\bar{c}_{{\ve r}, n,\sigma, \tau}{c}_{{\ve r},n+1,\sigma,\tau}+\bar{c}_{{\ve r}, n,\sigma, \tau}{c}_{{\ve r} +\Delta {\ve r} ,n,\sigma,\tau}+\text{H.c}\big)
\nonumber \\&&- \frac{J_k}{4} \int^\beta_0  d\tau \sum_{{\ve r},\sigma } \big( \bar{c}_{{\ve r},1, \sigma,\tau} {d}_{{\ve r},\sigma, \tau}+\text{H.c} \big)^2- \frac{J_h}{4}\int^\beta_0  d\tau \sum_{{\ve r},\sigma } \big( \bar{d}_{{\ve r},\sigma, \tau}  {d}_{{\ve r}+\Delta{\ve r},\sigma, \tau}+\text{H.c}\big)^2 \nonumber \\&&+\int^\beta_0  d \tau \sum_{{\ve r},\sigma} \bar{d}_{{\ve r},\sigma,\tau} (\partial_\tau+ i \lambda ({\ve r},\tau)) {d}_{{\ve r},\sigma,\tau}- \int^\beta_0  d\tau i \lambda({\ve r},\tau) \nonumber \nonumber \\&&-\Gamma  \int^\beta_0  d\tau \int^\beta_0 d \tau^\prime  \sum_{{\ve r}}  \Big(\frac{1}{2}\sum_{\sigma,\sigma^\prime} \bar{d}_{{\ve r},\sigma,\tau}  \ve{\sigma}_{\sigma,\sigma^\prime} {d}_{{\ve r},\sigma^\prime,\tau}\Big)\chi^0(\ve{0}, \tau-\tau^\prime)\Big(\frac{1}{2}\sum_{\sigma,\sigma^\prime} \bar{d}_{{\ve r},\sigma,\tau^\prime}  \ve{\sigma}_{\sigma,\sigma^\prime} {d}_{{\ve r},\sigma^\prime,\tau^\prime}\Big).
\end{eqnarray}
In the above  the summation  $\sum_{\r, n, \sigma}$  runs over the unit cell index $\r=1  \cdots  L$, the orbital index $n=1 \cdots  L$ and the  $z$ component of  spin $\sigma=\uparrow(\downarrow)$. Furthermore, the scalar Lagrange multiplier $\lambda ({\ve r},\tau)$ enforces the constraint and the  last term corresponds to the dissipation where $\Gamma$ is a dissipative coupling constant and $\chi^0(\ve{0}, \tau-\tau^\prime)\propto\frac{1}{(\tau-\tau^\prime)^2}$. Note that  in Eq.~(\ref{Sact}) the dissipation is added in  an  impromptu manner.

Next, we allow for  the Kondo screening, the spinon hopping and the magnetic ordering along the spin chain with the following  bond mean field decouplings,
\begin{eqnarray}
V({\ve r},\tau)&=&\sum_{\sigma}\big \langle  \bar{c}_{{\ve r},1,\sigma,\tau} {d}_{{\ve r}, \sigma,\tau}\big\rangle =\sum_\sigma\big\langle  \bar{d}_{{\ve r},\sigma,\tau} {c}_{{\ve r},1,\sigma,\tau} \big\rangle \\
\chi({\ve r},\tau)&=&\sum_\sigma\big\langle  \bar{d}_{\r,\sigma,\tau} {d}_{{\ve r}+\Delta{\ve r},\sigma,\tau}\big\rangle=\sum_\sigma \big\langle  \bar{d}_{{\ve r}+\Delta {\ve r},\sigma,\tau} {d}_{\ve{r},\sigma,\tau}\big\rangle\\
m({\ve r},\tau)&=&\frac{1}{2}  \sum_
 \sigma\big\langle  \sigma \bar{d}_{{\ve r}, \sigma,\tau} {d}_{{\ve r},\sigma,\tau} \big\rangle  e^{-i \Q.\r}=\frac{1}{2} \sum_\sigma  \big\langle \sigma \bar{d}_{{\ve r}, \sigma, \tau^\prime} {d}_{{\ve r},\sigma, \tau^\prime} \big\rangle  e^{-i \Q.\r}.
\end{eqnarray}
Here, $\Q$ is the antiferromagnetic wave vector,  $V(\ve{r},\tau)$ is  the hybridisation order parameter (or  Higgs  condensate)  between ${c}$ and ${d}$ electrons, $\chi (\ve{r},\tau)$  is the spinon hopping parameter along the spin chain and $m(\ve{r},\tau)$ denotes  the  staggered magnetic order parameter along the $z$-direction.

The action of Eq.~({\ref{Sact}}) in terms of  above mean-field order parameters  can be written as,
 \begin{eqnarray}\label{Sactmf}
{\mathcal S}(\bar{\phi}, \phi,V, \chi,m, \lambda) &=&\int^\beta_0 d \tau  \sum_{{\ve r},n,\sigma}\bar{c}_{{\ve r}, n,\sigma, \tau}\partial_\tau  {c}_{{\ve r},n,\sigma,\tau} -t \int^\beta_0 d \tau\sum_{{\ve r}, n,\sigma} \big(\bar{c}_{{\ve r}, n,\sigma, \tau}{c}_{{\ve r},n+1,\sigma,\tau}+\bar{c}_{{\ve r}, n,\sigma, \tau}{c}_{{\ve r} +\Delta {\ve r} ,n,\sigma,\tau}+\text{H.c}\big)
\nonumber \\&&- \frac{J_k}{2} \int^\beta_0  d\tau \sum_{{\ve r},\sigma} V({\ve r},\tau)\big( \bar{c}_{{\ve r},1, \sigma,\tau} {d}_{{\ve r},\sigma, \tau}+\text{H.c}\big)- \frac{J_h}{2}\int^\beta_0  d\tau \sum_{{\ve r}, \sigma}  \chi ({\ve r},\tau)\big( \bar{d}_{{\ve r},\sigma, \tau} {d}_{{\ve r}+\Delta {\ve r},\sigma, \tau}+\text{H.c}\big) \nonumber \\&&+\int^\beta_0  d \tau \sum_{\ve{r},\sigma} {\bar d}_{\ve{r},\sigma,\tau} \big(\partial_\tau+ i \lambda ({\ve r},\tau)\big) {d}_{{\ve r},\sigma,\tau}
- \frac{\alpha }{2}\times \int^\beta_0  d\tau  \sum_{{\ve r} ,\sigma} m({\ve r},\tau)  \big(\sigma\bar{d}_{{\ve r},\sigma, \tau}  {d}_{{\ve r}+\Q,\sigma, \tau}+\text{H.c}\big)\nonumber \\&&+\int^\beta_0  d \tau \sum_{{\ve r}} \Big(\frac{J_k}{2}  |V({\ve r},\tau)|^2+\frac{J_h}{2}  |\chi({\ve r},\tau)|^2+\alpha  |m({\ve r},\tau)|^2-i \lambda({\ve r},\tau)\Big).
\end{eqnarray}
In the above $\alpha$ denotes the renormalised dissipative coupling, $\alpha\propto  \Gamma \int d \tau\frac {1}{\tau^2}$.

 To find the saddle-point solutions,
 \begin{eqnarray}\label{Sadle}
\frac { d {\mathcal S}(\bar{ \phi}, \phi, V, \chi,m, \lambda)}{d V({\ve r},\tau)}=0, \quad  \frac { d {\mathcal S}(\bar{\phi}, \phi,V, \chi,m, \lambda)}{d \chi({\ve r},\tau)}=0, \quad
 \frac { d {\mathcal S}( \bar{\phi}, \phi, V, \chi,m, \lambda)}{d m({\ve r},\tau)}=0, \quad  \frac { d {\mathcal S}(\bar{\phi}, \phi,V, \chi,m, \lambda)}{d \lambda({\ve r},\tau)}=0
 \end{eqnarray}
we restrict  the   search to  space- and time-independent mean-field order parameters; $V({\ve r},\tau)=V$, $\chi({\ve r},\tau)=\chi$, $m({\ve r},\tau)=m$ and enforces the constraint  on average  $\lambda({\ve r},\tau)=  \lambda$.  Hence,  from  now onward we  will work in the   Hamiltonian formalism  with   static mean-field order parameters.

 The mean-field Hamiltonian can be written as,
 \begin{eqnarray}
\hat{H}_{mf}&=&-t\sum_{{\ve r}, n,\sigma} \big(\hat{c}^\dagger_{{\ve r}, n,\sigma} \hat{c}_{{\ve r},n+1,\sigma}+\hat{c}^\dagger_{{\ve r}, n,\sigma} \hat{c}_{{\ve r} +\Delta {\ve r} ,n,\sigma}+\text{H.c}\big)
 - \frac{J_kV}{2} \sum_{{\ve r},\sigma} \big( \hat{c}^{\dagger}_{{\ve r},1, \sigma}  \hat{d}_{{\ve r}, \sigma}+\text{H.c}\big) \nonumber \\&&- \frac{J_h\chi}{2} \sum_{{\ve r},\sigma} \big( \hat{d}^{\dagger}_{{\ve r}, \sigma}  \hat{d}_{{\ve r}+\Delta {\ve r}, \sigma}+\text{H.c}\big)+\lambda \sum_{{\ve r},\sigma} \hat{d}^\dagger_{{\ve r},\sigma} \hat{d}_{{\ve r},\sigma}  -\frac{\alpha m}{2} \sum_{{\ve r},\sigma} \big(\sigma \hat{d}^{\dagger}_{{\ve r}, \sigma}  \hat{d}_{{\ve r}+\Q, \sigma} +\text{H.c}\big)+N_ue_0
\end{eqnarray}
with, $e_0=(\frac{J_kV^2} {2}+\frac{J_h\chi^2} {2}+\alpha m^2-\lambda)$ and  $N_u$  is the number of unit cells. Hereafter, we set $\lambda=0$ for a particle-hole-symmetric band.

Further, using the Fourier transforms,
\begin{eqnarray}
\hat{c}_{{\ve r}, n, \sigma} =\frac{1}{ \sqrt{N_u/2}} \sum_{\k\in \mbox{MBZ}} e^{i \k. \ve{r}} \hat{c}_{k, n,\sigma}, \quad \quad \hat{d}_{\ve{r},\sigma} =\frac{1}{ \sqrt{N_u/2}} \sum_{\k\in \mbox{MBZ}} e^{i \k.\ve{r}} \hat{d}_{\k,\sigma}
\end{eqnarray}
the mean-field Hamiltonian in  momentum space  can be written as,
\begin{eqnarray}
\hat{H}_{mf}=\sum_{\k\in \mbox{MBZ},n,\sigma} \hat{\ve{\phi}}^\dagger_{\k,n,\sigma} M(\k)  \hat{\ve{ \phi}}_{\k,n,\sigma }+N_ue_0
\label{Hmfk_mag}
\end{eqnarray}
with  $\ve{\hat{\phi}}^\dagger_{\k,n,\sigma} =\left(\hat{c}^{\dagger}_{\k,1,\sigma},\hat{c}^{\dagger}_{\k+\Q,1,\sigma},\hat{c}^{\dagger}_{\k,2,\sigma},\hat{c}^{\dagger}_{\k+\Q,2,\sigma},\hat{c}^{\dagger}_{\k,3,\sigma},\hat{c}^{\dagger}_{\k+\Q,3,\sigma}, \cdots,  \hat{c}^{\dagger}_{\k,L,\sigma},\hat{c}^{\dagger}_{\k+\Q,L,\sigma}, \hat{d}^{\dagger}_{\k,\sigma} , \hat{d}^{\dagger}_{\k+\Q,\sigma}  \right) $  and
\begin{eqnarray*}
& & M(\k)=  \\
& & \left( \begin{array}{cccccccccccccccc}
-2t\cos \k &0&-t&0&0&0&\cdots &-t &0&\frac{-J_kV}{2}&0  \\
0&2t\cos \k&0 &-t&0&0&\cdots &0&-t &0&\frac{-J_kV}{2} \\
-t&0&-2t\cos \k &0&-t&0&\cdots &0&0 &0&0  \\
0&-t&0&2t\cos \k &0&-t&\cdots &0&0 &0&0  \\
0&0&-t&0&-2t\cos \k &0&\cdots&0&0 &0&0  \\
0&0&0&-t&0 &2t\cos \k &\cdots &0&0 &0&0  \\
0&0&0&0&-t &0&\cdots&0&0 &0&0  \\
\cdots&\cdots&\cdots&\cdots&\cdots  &\cdots & \cdots & \cdots  & \cdots&\cdots &\cdots \\
-t&0& 0&0& 0&0&\cdots  &-2t\cos \k &0 & 0&0 \\
0&-t&0&0&0&0&\cdots &0&2t\cos \k & 0&0  \\
\frac{-J_kV}{2}&0&0&0&0&0&\cdots &0&0 &- J_h \chi\cos \k & -\frac{\alpha m\sigma}{2} \\
0&\frac{-J_kV}{2}&0&0&0&0&\cdots &0 &0& -\frac{\alpha m\sigma}{2}& J_h \chi\cos \k \\
\end{array}
\right).%
\end{eqnarray*}
Diagonalizing the mean-field Hamiltonian, $U^{\dagger}(\k) M (\k)  U(\k)  =  \text{Diag} \left(  E_{\k,1},   \cdots,  E_{\k,2(L + 1)} \right)$ gives:
\begin{eqnarray}
\hat{H}_{mf}=N_ue_0+\sum_{\k\in\mbox{MBZ},n,\sigma} E_{\k,n,\sigma} {\hat{\gamma}}^\dagger_{\k,n,\sigma} {\hat{\gamma}}_{\k,n,\sigma}.
\end{eqnarray}
 Here,  the $\k$ summation goes over the reduced magnetic Brillouin zone (MBZ) and  $\ve{\hat{\gamma}}^\dagger_{\k} = \ve{\hat{\phi}}^\dagger_{\k} U(\k)$.

\begin{figure}[htbp]
\centering
\includegraphics[width=0.5\textwidth]{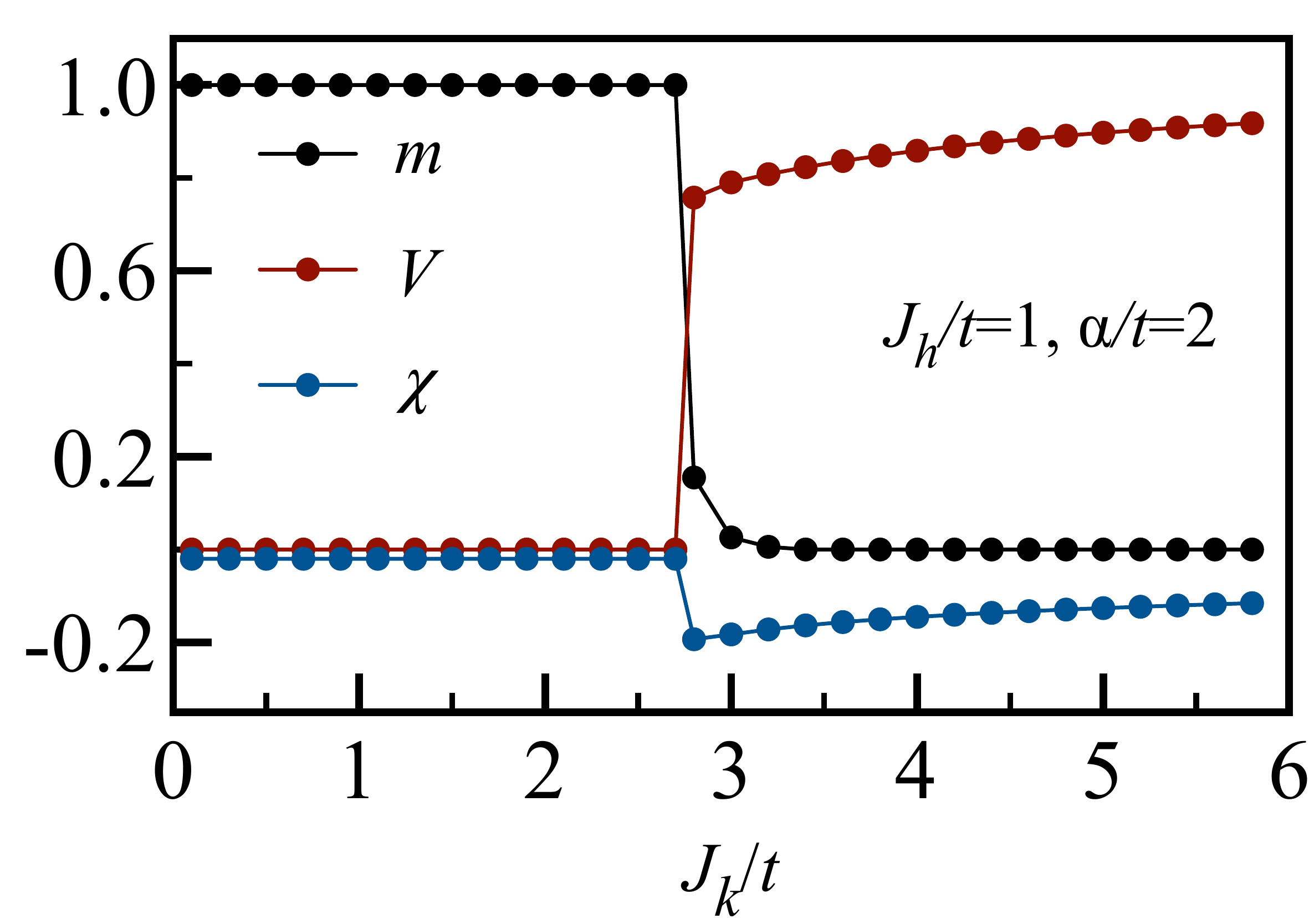}
\caption{Mean-field order parameters as a function of  $J_k/t$ at $N_u=L_x=L_y=L=100$.}
\label{mnfld_prmtr_vs_Jhalpha}
\end{figure}

Fig.~\ref{mnfld_prmtr_vs_Jhalpha}  plots  the mean-field order parameters $V$, $\chi$, and $m$ as a function of Kondo coupling.  Specifically, by fixing $\alpha/t=2$ and $J_h/t=1$, the mean-field theory gives the two phases,  (i) a dissipation-induced magnetically ordered phase  characterized  by $m=1, V=0, \chi=0$  and (ii) a disordered Kondo-screened phase  characterized  by  $m=0, V\ne0, \chi\ne0$.

Further, to explore heavy-Fermi-liquid properties at large $J_k$ limit we set $\alpha=0$ in Eq.~(\ref{Hmfk_mag}) and  self consistently search for mean-field order parameters $V$ and $\chi$ in the Brillouin zone (BZ).  In this case for $J_k/t<2$ we  observe  a Kondo-breakdown phase  and  for $J_k/t\gtrsim2$ a transition into the  Kondo-screened phase.  Considering that the Kondo-breakdown phase is an artefact of the mean-field approach in the following we only focus on the Kondo-screened paramagnetic phase.

\begin{figure}[htbp]
\centering
\includegraphics[width=0.7\textwidth]{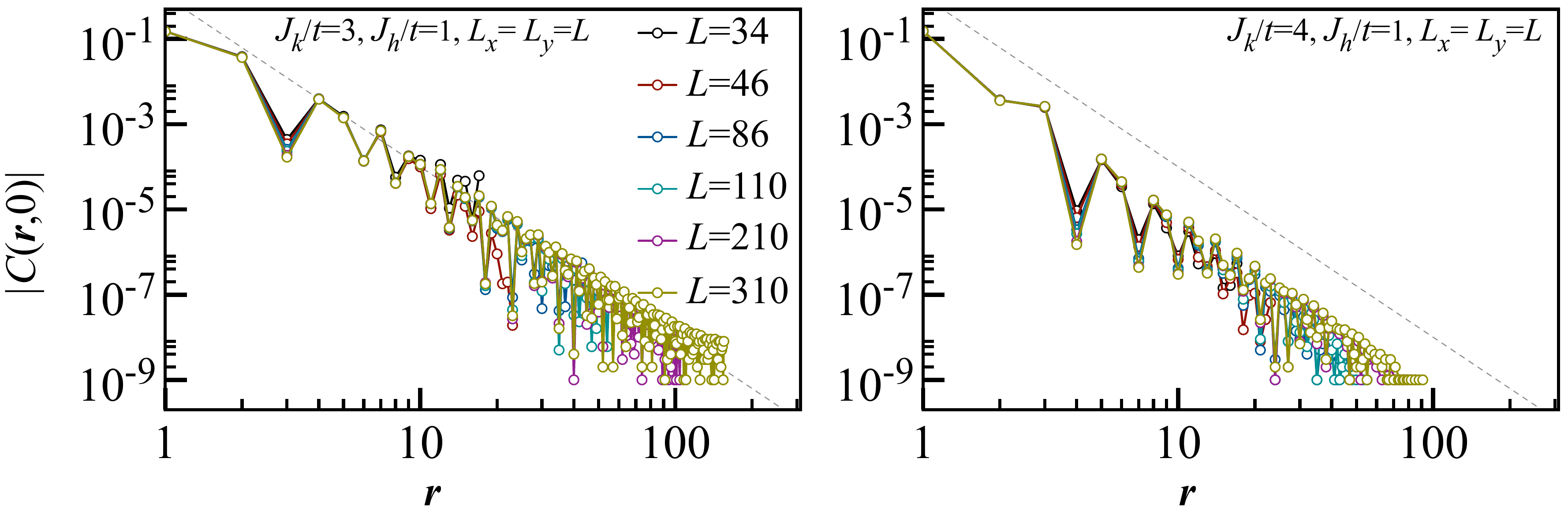}%
\caption{Equal-time spin-spin correlation function $ |C(\r,0)|$ along the spin chain  with respect to distance $r$ in the Kondo-screened phase  within the  mean-field calculation. The dashed grey line represents the  $1/r^4$ power  law.} 
 \label{LogCr_vs_r_LargeN}

 ~

 \includegraphics[width=0.7\textwidth]{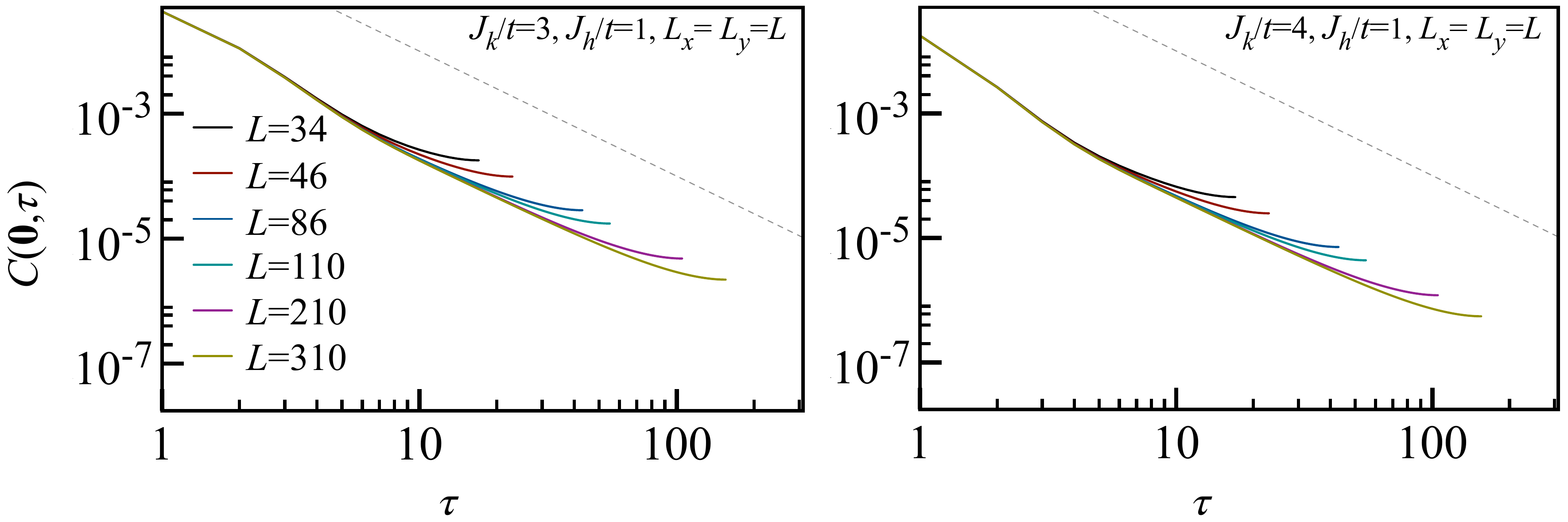}%
\caption{Time-displaced spin-spin correlation function $C(\ve{0},\tau)$ along the spin chain  with respect to imaginary $\tau$ in the Kondo-screened phase within the  mean-field calculation. The dashed grey line represents the  $1/\tau^2$ power  law.} 
 \label{LogCtau_vs_tau_LargeN}
\end{figure}

Fig.~\ref{LogCr_vs_r_LargeN} plots  the  mean-field results for equal-time spin-spin correlation function $C(\r,0)=\frac{1}{L} \sum_\q e^{i(\q+\Q).\r}\big\langle \hat{S}^z(\q, 0) \hat{S}^z(-\q, 0) \big\rangle$  along the spin chain as a function of $r$ at $J_k/t=3$ and $J_k/t=4$. Here, the spins acquire the $1/r^4$ power law decay of 2D conduction electrons. Correspondingly, Fig.~\ref{LogCtau_vs_tau_LargeN} plots the time displaced spin-spin correlation function $C(\ve{0},\tau)=\frac{1}{L} \sum_\q \big\langle T_\tau \hat{S}^z(\q, 0) \hat{S}^z(-\q, \tau) \big\rangle$  as a function of imaginary time $\tau$.  Here, the  spins acquire the $1/\tau^2$ power-law decay of 2D conduction electrons.
%

\begin{figure}[htbp]
\centering
\includegraphics[width=0.7\textwidth]{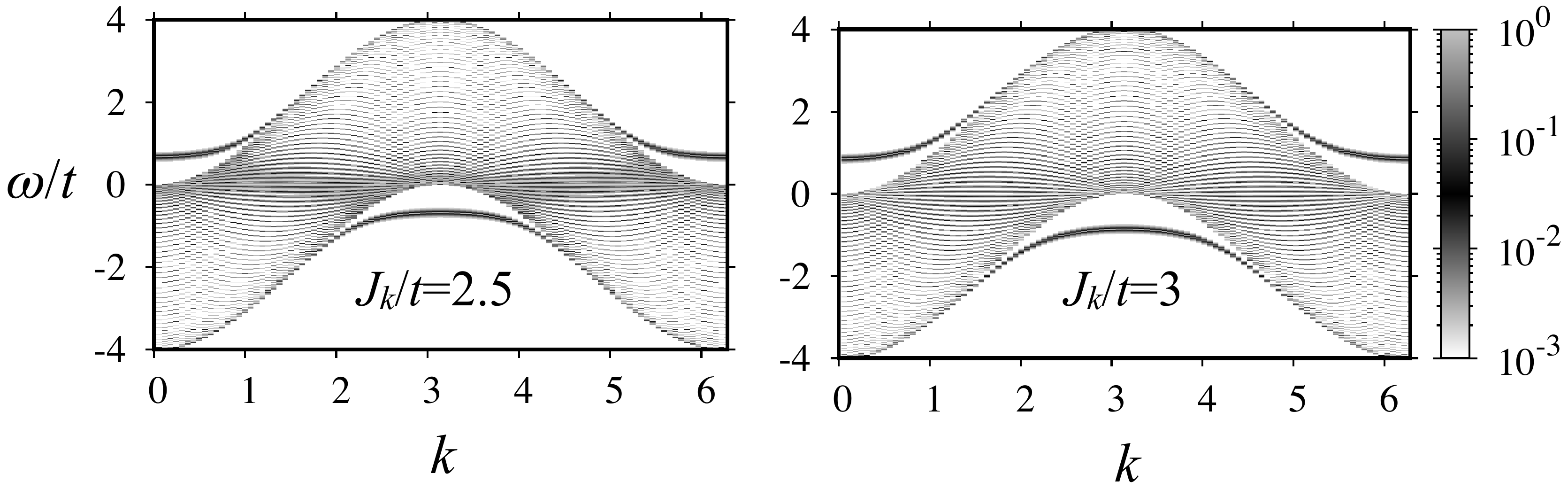}%
\caption{ The $d$ spectral function  $A_d(k,\omega)$ as a function of energy $\omega/t$ and momentum $k$ at $J_h/t=1$ obtained within the mean-field calculation.}
 \label{spectral_wieght_d}

 ~

 \includegraphics[width=0.7\textwidth]{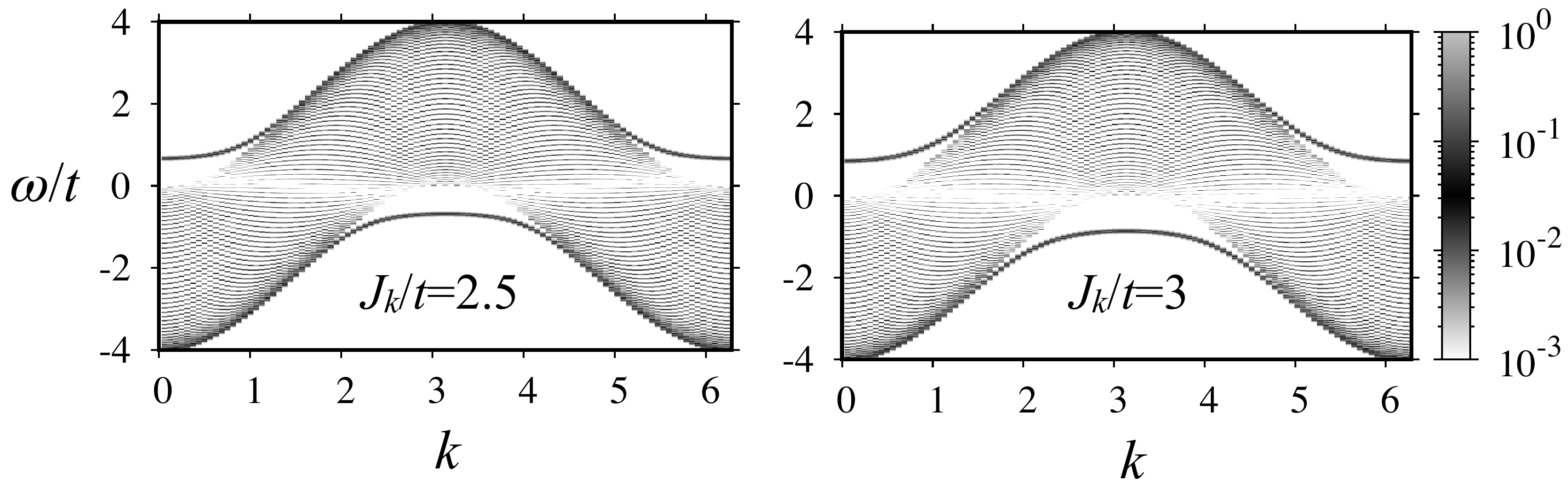}%
\caption{ The $c$ spectral function $A_c(k,\omega)$ as a function of energy $\omega/t$ and momentum $k$  at $J_h/t=1$ obtained within the  mean-field calculation.}
 \label{spectral_wieght_c}
 \end{figure}

To understand the  heavy  band  in the composite-fermion spectral function $A_\psi(k,\omega)$ (see  Figs.~7 (a) and (b) of the main paper) we compute the retarded Green functions $G^{\mbox{ret}}_d (\k, \omega)$ and $G^{\mbox{ret}}_c (\k, \omega)$ on $d$ electrons and the $c$ electrons  respectively as given below,
\begin{eqnarray}
G^{\mbox{ret}}_d (\k, \omega)=-i \int^{\infty}_0 d t e^{i \omega t}\sum_\sigma \big\langle  \big\{ \hat{d}_{\k, \sigma} (t),\hat{d}^\dagger_{\k, \sigma}(0)\big\}\big\rangle=  \sum_n \frac{\big|U_{\k, L+1,n}\big|^2}{\omega-E_{\k,n}+i0^+} \\
G^{\mbox{ret}}_c (\k, \omega)=-i \int^{\infty}_0 d t e^{i \omega t}\sum_\sigma \big\langle  \big\{ \hat{c}_{\k,1, \sigma} (t),\hat{c}^\dagger_{\k, 1,\sigma}(0)\big\}\big\rangle=\sum_n \frac{\big|U_{\k, 1,n}\big|^2}{\omega-E_{\k,n}+i0^+}.
\end{eqnarray}
Fig.~\ref{spectral_wieght_d} plots the $d$-spectral function $A_d(\k,\omega)=-\frac{1}{\pi} \mbox{Im} G^{\mbox{ret}}_d (\k,\omega)$ as a function of energy and momentum in the Kondo-screened phase.  Similarly,  Fig.~\ref{spectral_wieght_c} plots the spectral function of the Kondo-coupled row of conduction electrons $A_c(\k,\omega)=-\frac{1}{\pi} \mbox{Im} G^{\mbox{ret}}_c (\k,\omega)$.
As  apparent,  in  $A_d(\k,\omega)$,    we  observe  a  band  that detaches from the continuum   and  that is  reminiscent of the  the   heavy  fermion band  observed  within a  one-dimensional Kondo insulator   \cite{Tsunetsugu1997}.  This  heavy-fermion state  hybridizes with the other conduction electrons.   The  notion of  heavy Fermi liquid  for this  dimensional-mismatch  Kondo  problem  becomes  precise  when  considering the volume of the Fermi surface as described by Luttinger's theorem,  discussed in the next  section.


\section{Luttinger theorem for a hybrid-dimensionality Kondo lattice model}

In Ref.~\onlinecite{Oshikawa00a}, Oshikawa showed that if the low-energy theory of a standard Kondo lattice model is a conventional Fermi liquid, then the volume of the Fermi surface equals the sum of the density of the conduction electrons and density of the local moments (mod 2), i.e., the Fermi surface is `large'. Here we discuss Oshikawa's argument in the context of our model, or more generally, in the context of
models where the conduction electrons and the local moments have a dimensional mismatch. This allows one to give a precise meaning to the heavy Fermi liquid phase in such models.

Consider our model defined in Eq.~(1) of the main paper with local moments that are located along a chain at $y = 0$. We put this system on a $L_x \times L_y$ cylinder with periodic boundary condition along the $x$ direction, and denote by $\nu_{\sigma}$ the density of conduction electrons with spin $\sigma = \uparrow, \downarrow$, i.e., $\nu_\sigma = \langle \sum_i c^{\dagger}_{i \sigma} c^{}_{i\sigma}\rangle/(L_x L_y)$.  This system can be thought of as a translationally invariant one-dimensional system where each unit cell contains a single local moment with spin-1/2, and  $L_y \nu_\sigma $ conduction electrons with spin $\sigma$. We now adiabatically insert a $2 \pi$ flux of a gauge field that only couples to say, up-spin electrons. Let us denote the original Hamiltonian as $H(\Phi = 0)$ and the final Hamiltonian as $H(\Phi = 2\pi)$. If the original ground state $|\psi(\Phi = 0)\rangle$ has lattice momentum $p_{0x}$, and eigenenergy $E_0$, then after this adiabatic evolution, it will evolve to a new state $|\psi(\Phi = 2\pi)\rangle$  with the same lattice momentum $p_{0x}$ and,  same eigenenergy $E_0$ i.e.
$H(\Phi = 2 \pi)  |\psi(\Phi = 2\pi) \rangle  = E_0 |\psi(\Phi = 2\pi) \rangle$. However, $H(0) \neq H(2\pi)$ and in fact $U_\uparrow^{\dagger} H(\Phi = 2\pi) U_\uparrow = H(\Phi = 0)$ where $U_\uparrow = e^{\frac{2\pi i}{L_x} \sum_x x \left( n_\uparrow(x) + S^z(x)\right)}$ and $n_\uparrow(x) = \sum_{y = 1}^{L_y} c^{\dagger}_{x,y,\uparrow} c_{x,y,\uparrow}$. Now one may use the commutation relation between the translational operator $T_x$ and $U_\uparrow$ to show that the state $U_\uparrow|\psi(\Phi = 2\pi)\rangle$ carries momentum $p_{0x} + 2 \pi \left(\nu_\uparrow L_y + 1/2 \right)$ where we have assumed that the  magnetization of the local moments is zero. If the low-energy theory is a Fermi liquid, the insertion of $2 \pi$ flux shifts the Fermi surface of the up-spin quasiparticles by momentum $2 \pi/L_x$, and therefore, the momentum transferred also equals $\frac{2 \pi}{L_x}\times N^{L_x}_\uparrow$ where $N^{L_x}_\uparrow$ is the number of occupied momentum modes for the up-spin quasiparticles. Equating the two expressions for the momentum transferred, one finds $N^{L_x}_\uparrow = (\nu_\uparrow L_y + 1/2) L_x$. If the system did not have any local moments, then this same procedure would instead yield $N^{L_x}_\uparrow = \nu_\uparrow L_y L_x$. The additional `+1/2' inside the brackets in the expression for $N^{L_x}_\uparrow$ provides a precise meaning to the statement that the local moments are absorbed in the Fermi volume. One may run the same argument for the down-spin electrons, and obtain $N^{L_x}_\uparrow + N^{L_x}_\downarrow = (\nu L_y + 1) L_x$ where $\nu = \nu_\uparrow + \nu_\downarrow$ is the total density of conduction electrons. More generally, this equation provides a non-perturbative definition of the heavy Fermi liquid state for a mixed-dimensionality model such as ours.

\section{ Spin-1/2  XXZ  CHAIN on a 2D metal}

The Hamiltonian for a spin-1/2  XXZ chain on a 2D metal  can be written as
\begin{eqnarray}
 \Hhat &= &-t\sum_{\langle \i,\j\rangle}\big(\hat{\ve{c}}^{\dagger}_{\i} \hat{\ve{c}}^{}_{\j}+\text{H.c}\big)
 +\frac{J_k}{2}\sum^L_{\ve{r} =1}  \hat{\ve{c}}^{\dagger}_{\ve{r}}  \ve{\sigma} \hat{\ve{c}}^{}_{\ve{r}}
 \cdot \hat{\ve{S}}_{\ve{r}}+\sum^{L}_{\ve{r}=1}\left( \frac{J^{\perp}}{2}\big(\hat{S}^+_{\ve{r}} \hat{S}^-_{\ve{r}+\Delta\ve{r}}+\hat{S}^-_{\ve{r}} \hat{S}^+_{\ve{r}+\Delta\ve{r}}\big)+J^z \hat{S}^z_{\ve{r}} \hat{S}^z_{\ve{r}+\Delta\ve{r}}\right)
\label{model_ham_xxz}
\end{eqnarray}
 where $J^{\perp}$ and $J^z$ are the transverse and longitudinal exchange couplings along the spin chain.

The space and time displaced  spin-spin correlation power-law decays along the decoupled XXZ spin chain  are set by the Luttinger parameter $K=\Big[ \frac{2}{\pi} \cos^{-1} \big(-J^z/J^\perp\big) \Big]^{-1}$~\cite{Peschel75}. Specifically, the transverse and longitudinal  spin-spin correlation functions  are given by the power-law decays
\begin{eqnarray}
\chi^\perp(\r,\tau)=e^{i \Q.\r}\big\langle \hat{S}^+(\r,\tau) \hat{S}^-(0,0)  \big \rangle\propto \frac{1} {\sqrt{(v_s\tau)^2+r^2}^{\, 1/2K}}\label{xxz_cor_xx}\\
\chi^z(\r,\tau)=e^{i \Q.\r}\big\langle \hat{S}^z(\r,\tau) \hat{S}^z(0,0)  \big \rangle\propto \frac{1} {\sqrt{(v_s\tau)^2+r^2}^{\, 2K}}
\label{xxz_cor_zz}
\end{eqnarray}
with $v_s$ being the spin velocity. The case $K=1/2$ corresponds to the isotropic Heisenberg point $J^\perp=J^z$  where the power-law decay takes the form $1/\sqrt{(v_s\tau)^2+r^2}$. The regime $K>1/2$  corresponds to an anisotropic XXZ chain  with   $J^\perp>J^z$.

In power counting for $K>1/2$   the Kondo coupling is relevant and irrelevant with respect to the scaling dimensions of transverse and longitudinal spin components. Hence one can expect  to see a   stronger  tendency  towards dissipation-induced  ordering  in the  transverse correlations  along the XXZ chain.  To check this point of view  we have performed our QMC simulation at  $J^\perp=4 J^z$.  This choice of   anisotropy is motivated  by Co adatoms on Cu$_2$N/Cu(100) surface~\cite{Toskovic2016}.

\begin{figure}[h]
\centering
\includegraphics[width=0.45\textwidth]{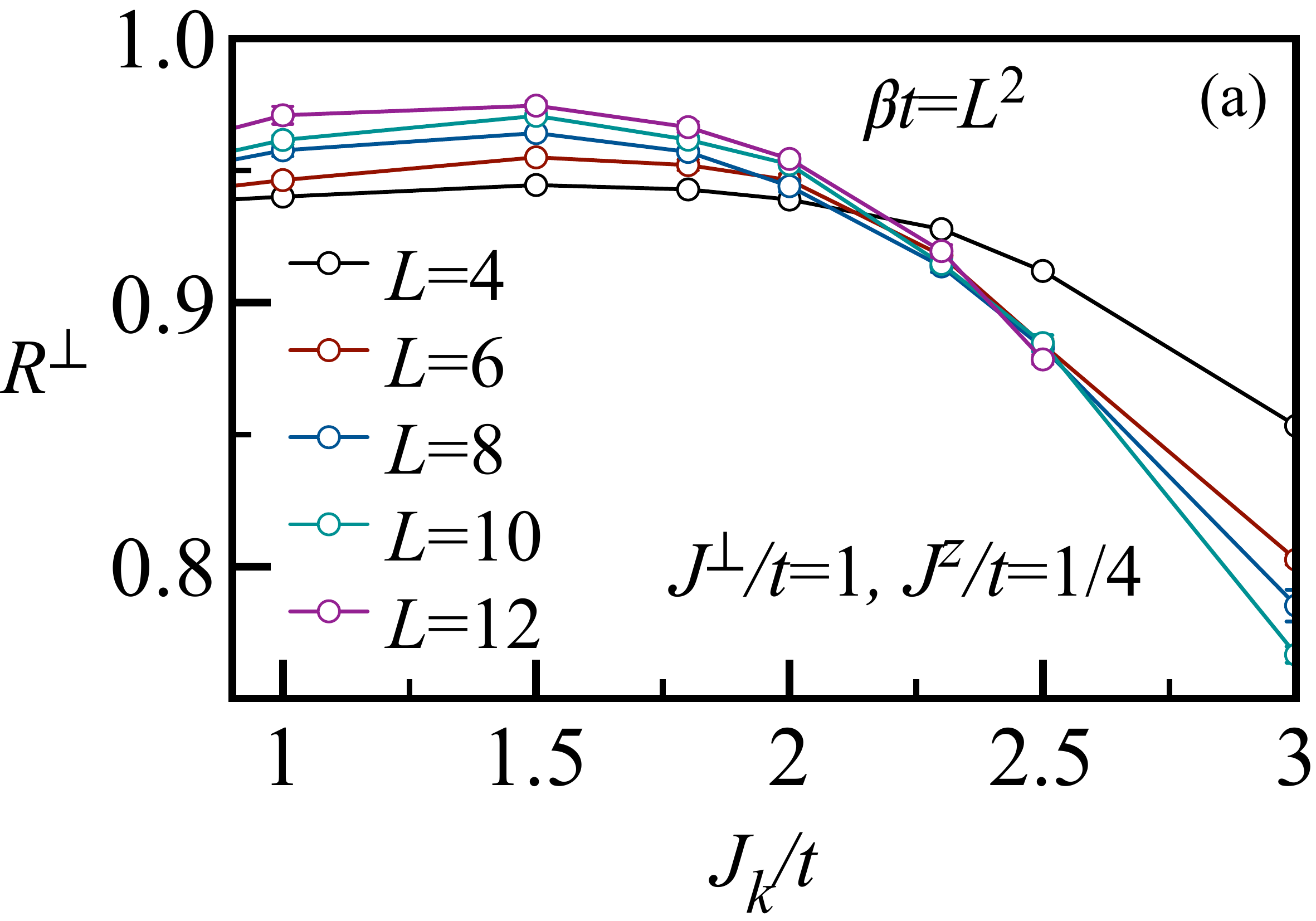}
\includegraphics[width=0.45\textwidth]{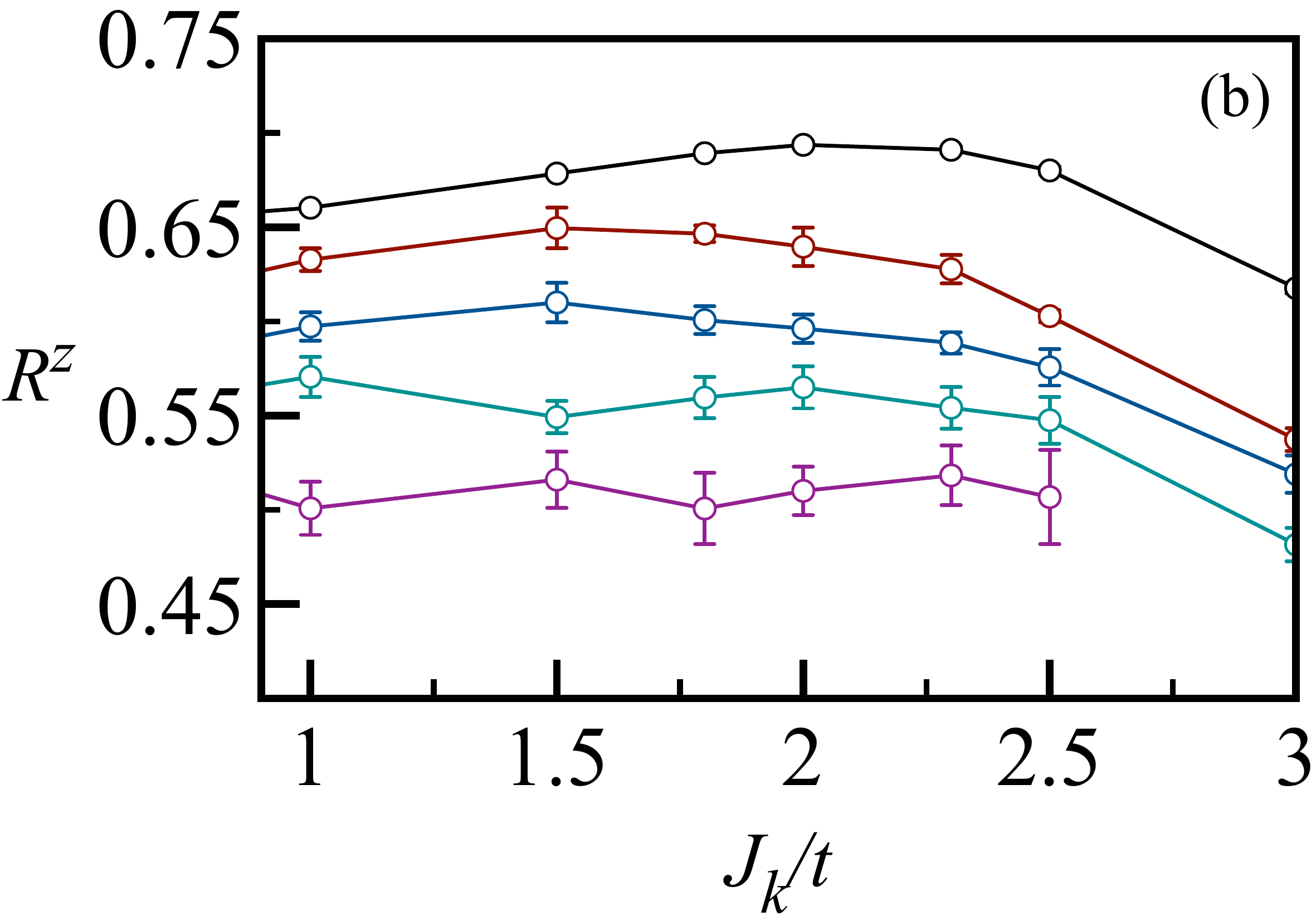}
\caption{QMC results for a spin-1/2 XXZ chain on a 2D metal at $J^\perp=4J^z$. Here, the Luttinger parameter reads $K=0.861429$. Left: Transverse component of  correlation ratio $R^\perp$ as a function of  $J_k/t$ at $\beta t=L^2$. Right: Longitudinal  component of  correlation ratio  $R^z$ as a function of  $J_k/t$ at $\beta t=L^2$. }
\label{Rperpz_XXZ_vs_Jk}
\end{figure}

Specifically, we focus on the transverse $R^{\perp}$ and longitudinal  $R^{z}$ components of the correlation ratio,
 \begin{eqnarray}
R^{\perp}=1-\frac{\chi^{\perp}{(\ve{Q}-\delta \ve{ k} },0 )} {\chi^{\perp}(\ve{Q},0)}, \quad \quad R^{z}=1-\frac{\chi^{z}{(\ve{Q}-\delta \ve{ k} },0 )} {\chi^{z}(\ve{Q},0)}.
\end{eqnarray}
Fig.~\ref{Rperpz_XXZ_vs_Jk} (a) and (b) plots  $R^{\perp}$   and $R^{z}$ as a function of $J_k/t$.  Clearly as a function of $J_k/t$ the transverse correlation ratio $R^\perp$ reveals a  quantum critical point at  $J^c_k/t\sim2.3$.  Noticeably, the enhancement in $R^\perp$ with increasing  $L$ is more pronounced for $J_k\lesssim2$ as compared to Fig.~2(b) of the main paper.  In contrast,  the longitudinal part $R^z$  does not show any critical behavior.
\section { FURTHER QMC RESULTS}
\label{QMC_support}

\begin{figure}[htbp]
\centering
\includegraphics[width=0.8\textwidth]{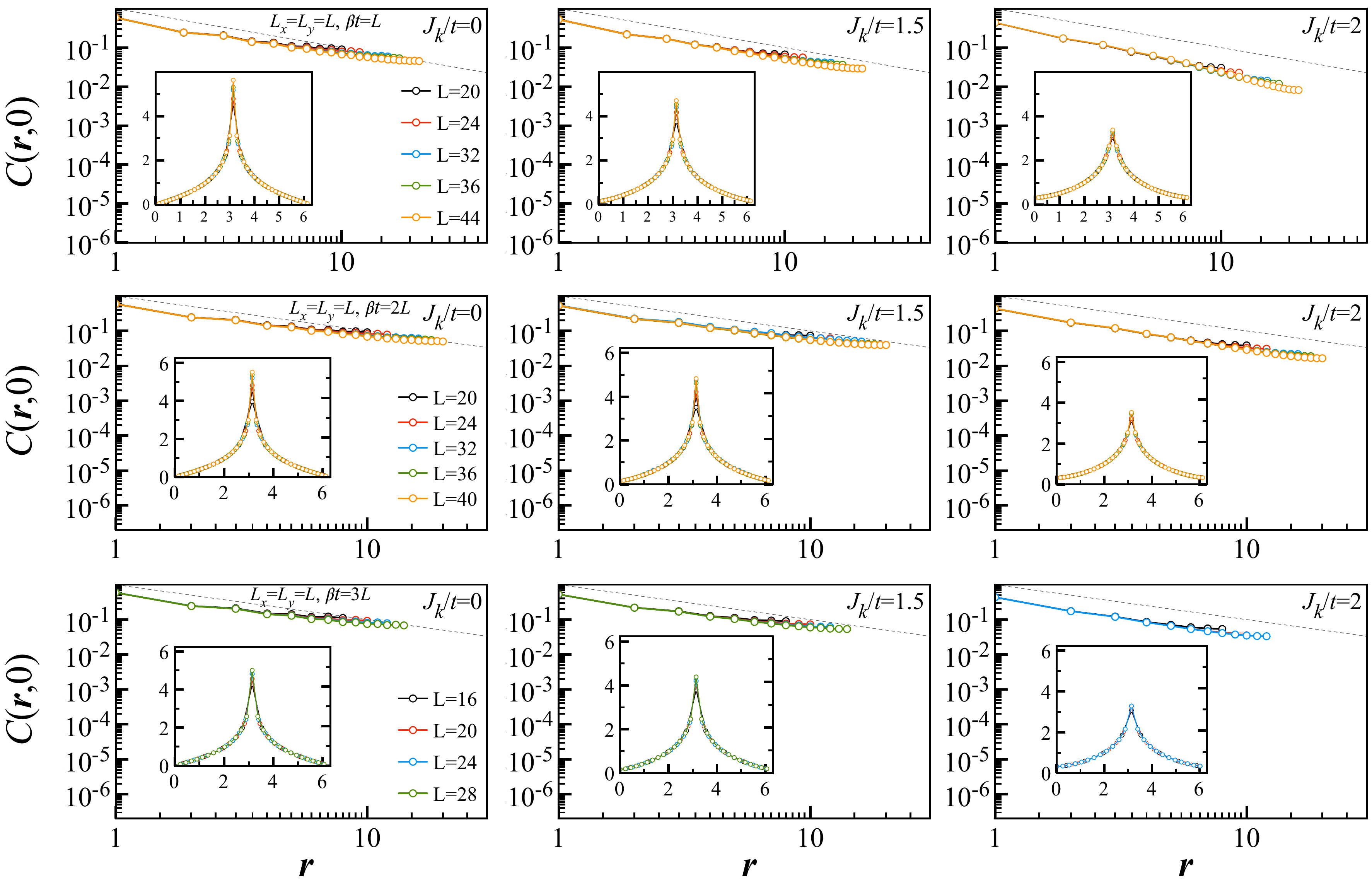}
\caption{Same as Fig.~4 (a)-(c) of the main paper on larger system sizes at  $\beta t =L, 2L, 3L$ in the ordered phase. Equal-time spin-spin correlation function $C(\r, 0)$  with respect to distance $r$ at $J_h/t=1$ for given $J_k/t$  values. Top row,  $\beta t=L$. Middle row, $\beta t=2L$.  Bottom row, $\beta t=3L$. The dashed grey line denotes the $1/r$  power  law. The insets plots corresponding  static spin structure factor $S(k)$ with respect to momentum vector $k$. }
\label{LogSrvslograll123betaLLxeqLy}

~

\includegraphics[width=0.8\textwidth]{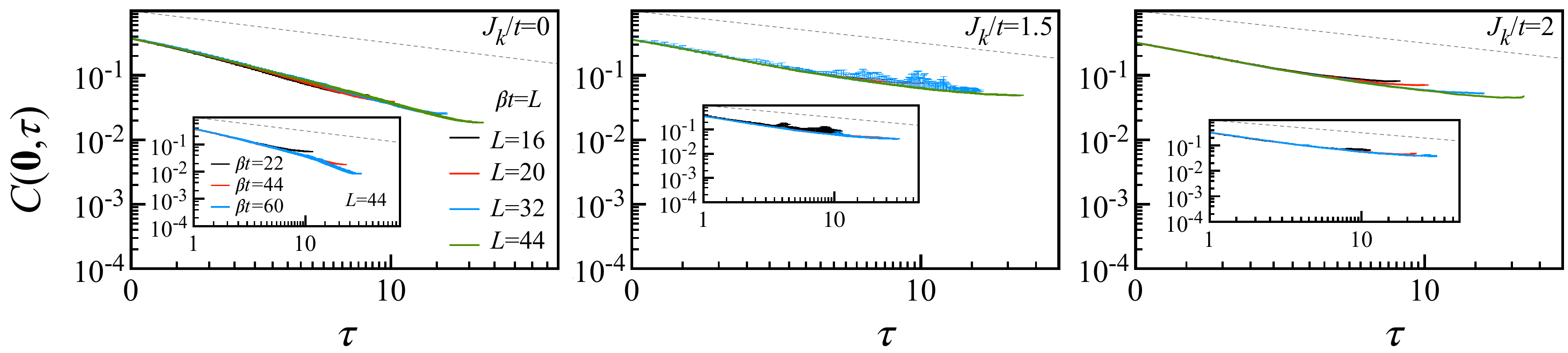}
\caption{Same as Fig.~4 (b)-(d) of the main paper on larger system sizes at $\beta t=L$ in the ordered phase. Time-displaced spin-spin correlation function $C(\ve{0},\tau)$  with respect to  imaginary  time $\tau$ at $\beta t=L, J_h/t=1$. Left,  $J_k/t=0$. Middle,  $J_k/t=1.5$.  Right,  $J_k/t=2$. The dashed grey line denotes the $1/\sqrt{\tau}$  power  law. The insets plot the same at a fixed $L=44$ for given inverse temperatures $\beta t$.}
\label{Stau_vs_tau_betaL_L44}
\end{figure}

In this section we present  supplemental   QMC results   supporting the  interpretation  in the main paper.
In  Fig.~\ref{LogSrvslograll123betaLLxeqLy}    we  present the  equal  time   spin-spin correlations    as  a function of  size  and  temperature.  In the  Heisenberg  limit,  $J_k=0$,   the  dynamical  exponent reads  $z=1$  such   that  system  sizes  $\beta    \propto   L  $   suffice   to access  ground-state properties.  Upon  inspection we  see   practically no  difference  in the  data  when  considering   $\beta t   =   L$,  $\beta t   =  2 L  $,  and $\beta  t  =  3 L  $.   In the  vicinity of the quantum phase  transition,   $J_k /t  = 2$,      we  see  that since  $z \simeq  2$  the   choice $\beta t= L $    results  in  high  temperature  data.  In  particular   at  $\beta t= L $  we  see  that the spin-spin correlations decay   quicker  than   $1/r$   and   we   ultimately  expect   to see an  exponential  decay in the large-size limit. This  exponential  decay  stems from thermal  fluctuations.

In  Fig.~\ref{Stau_vs_tau_betaL_L44}   we  consider  the  imaginary   decay.   Consider  the $L=44$  lattice  as  a  function of  temperature,  as  shown in the   insets.  Lowering the temperature  for the Heisenberg  case  shows that one  quickly  resolves    the  finite  size  gap   given  by
$1/L$. Beyond  this scale, the  imaginary-time data  falls  off  exponentially.   In contrast  in the vicinity  of the  critical point  as  well  as  in  the crossover   dissipative phase,  we  do not  seem to  be able to resolve  the finite-size  gap.   In fact,  in the  vicinity of the  critical point,  we  expect    it  to  scale  as
$1/L^z$  with  $z=2$.

\begin{figure}[htbp]
\centering
\includegraphics[width=0.8\textwidth] {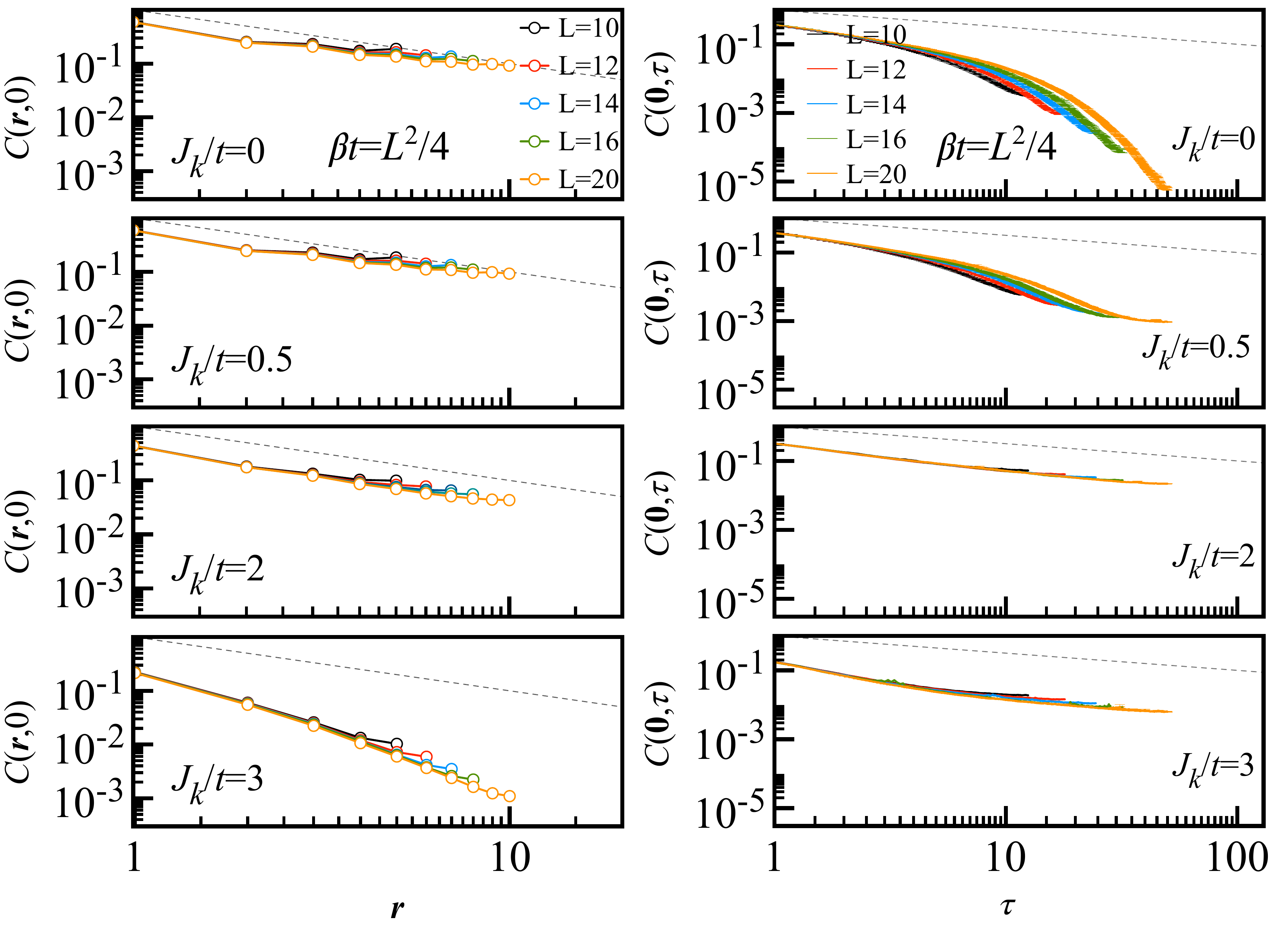}
\caption{Same as Fig.~4 of the main paper at  $\beta t =L^2/4$ and $J_h/t=1$ for given $J_k/t$ values. Left: Equal-time correlations $C(\ve{r},0)$ with respect to distance $r$.   Here, the dashed grey lines denote the $1/r$ power law.  Right: Time-displaced correlations  $C(\ve{0},\tau)$ with respect to imaginary time $\tau$. Here, the dashed grey lines denote the $1/\sqrt{\tau}$ power law.}
 \label{LogCr_Stau_vs_r_tau_betaLsqby4}
 
 ~
 
\includegraphics[width=0.8\textwidth] {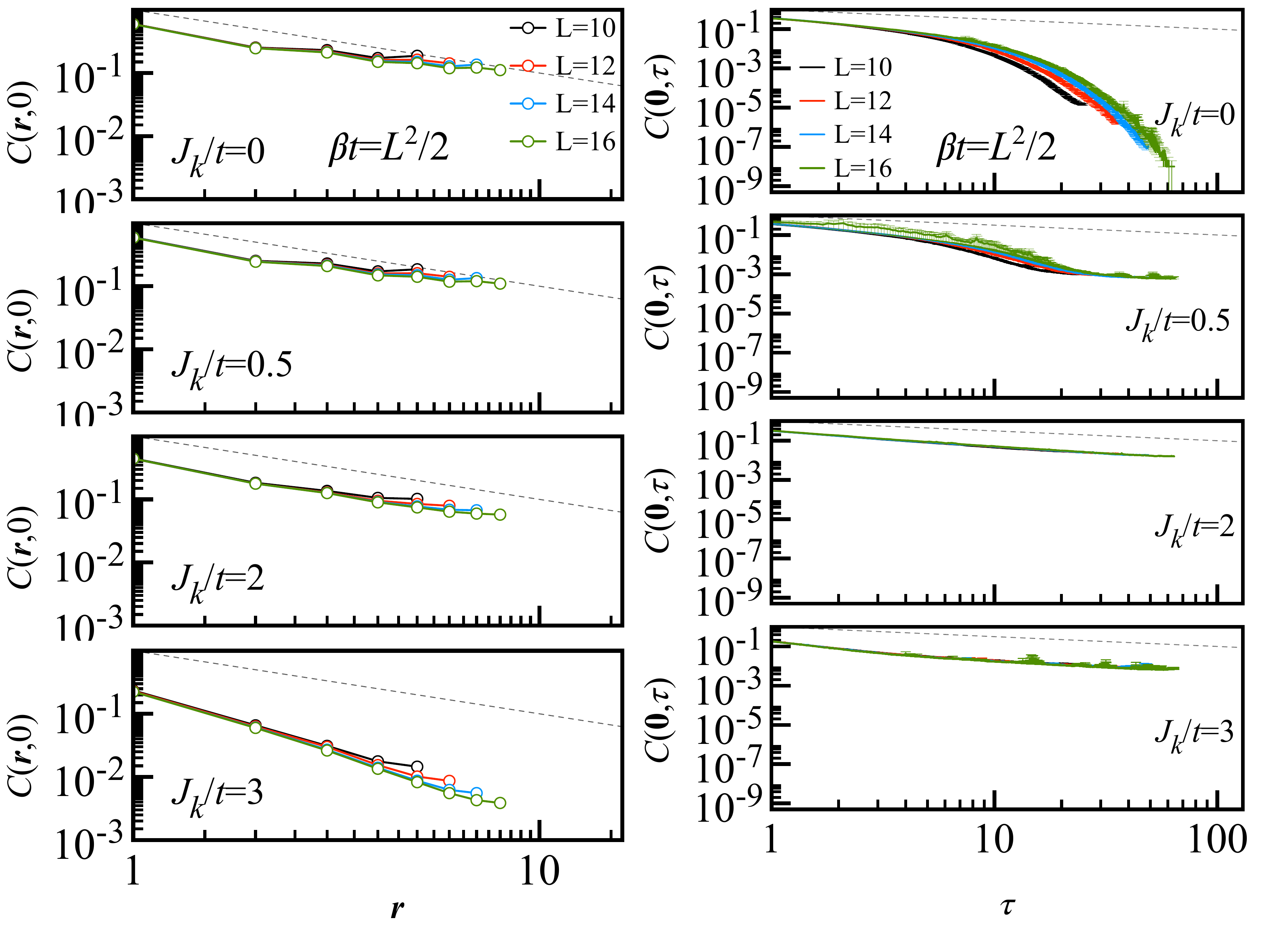}
\caption{Same as Fig.~4 of the main paper at  $\beta t =L^2/2$ and $J_h/t=1$ for given $J_k/t$ values. Left: Equal-time correlations $C(\ve{r},0)$ with respect to distance $r$.  Here, the dashed grey lines denote the $1/r$ power law.  Right: Time-displaced correlations  $C(\ve{0},\tau)$ with respect to imaginary time $\tau$. Here, the dashed grey lines denote the $1/\sqrt{\tau}$.}
 \label{LogCr_Stau_vs_r_tau_betaLsqby2}
\end{figure}

\begin{figure}[h]
 \includegraphics[width=0.32\textwidth]{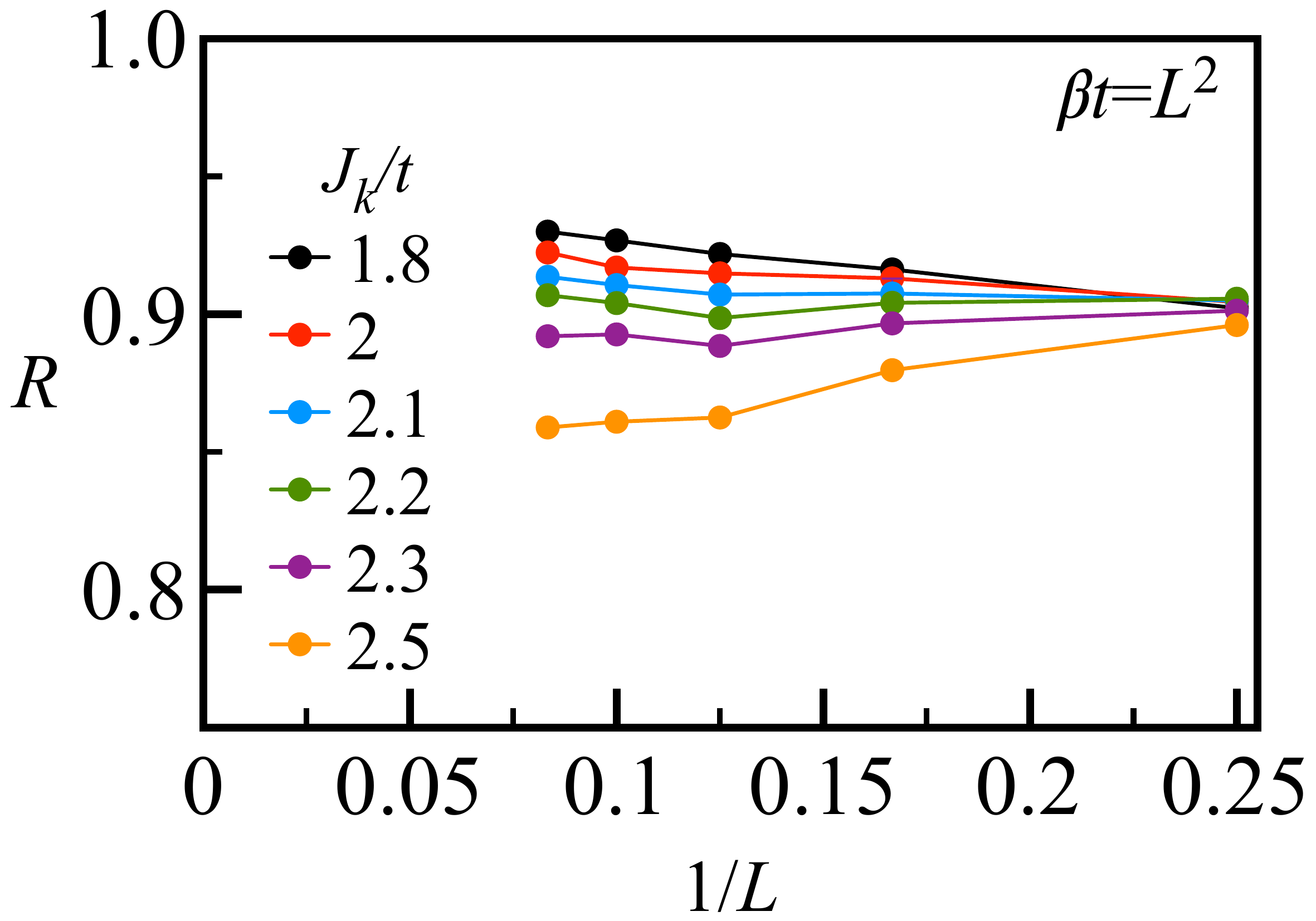}
\includegraphics[width=0.32\textwidth]{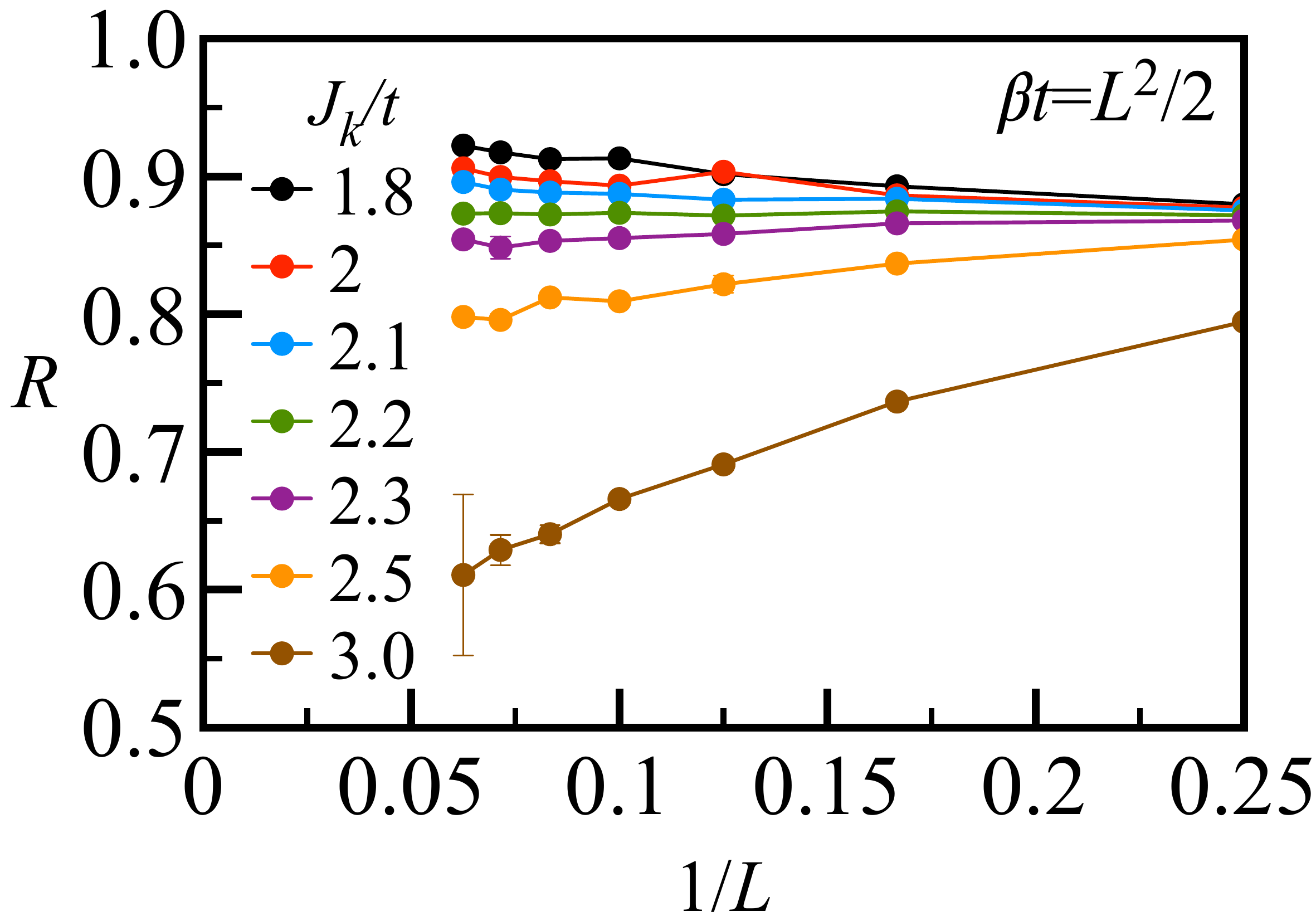}
\includegraphics[width=0.32\textwidth]{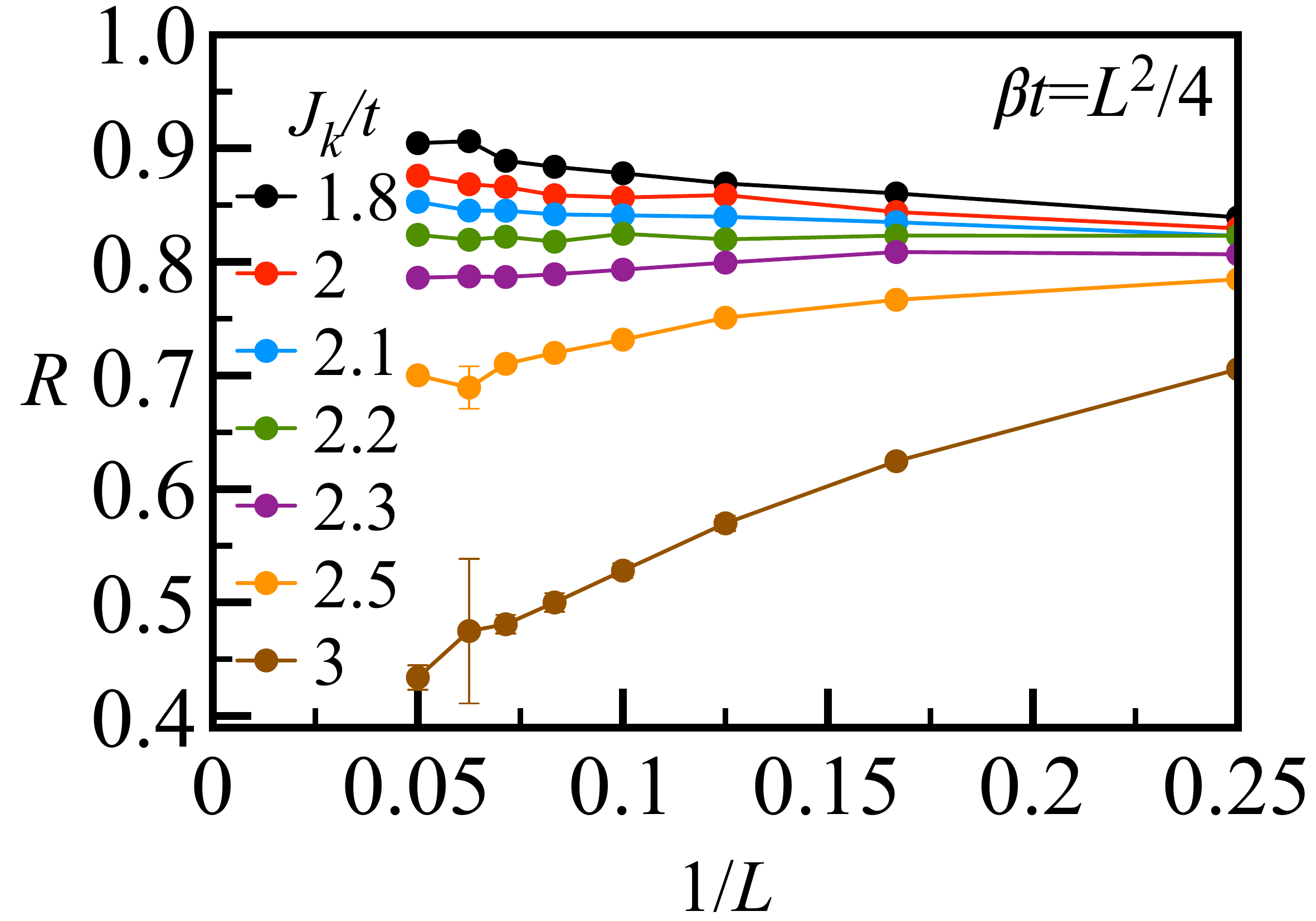}
\caption{Left: Correlation ratio $R$ as a function of  $1/L$ at  $\beta t=L^2, J_h/t=1$ for given $J_k/t$ values. Middle: Same at  $\beta t=L^2/2$. Right: Same at $\beta t=L^2/4$.} %
\label{R_vs_1byL}

~

\includegraphics[width=0.45\textwidth]{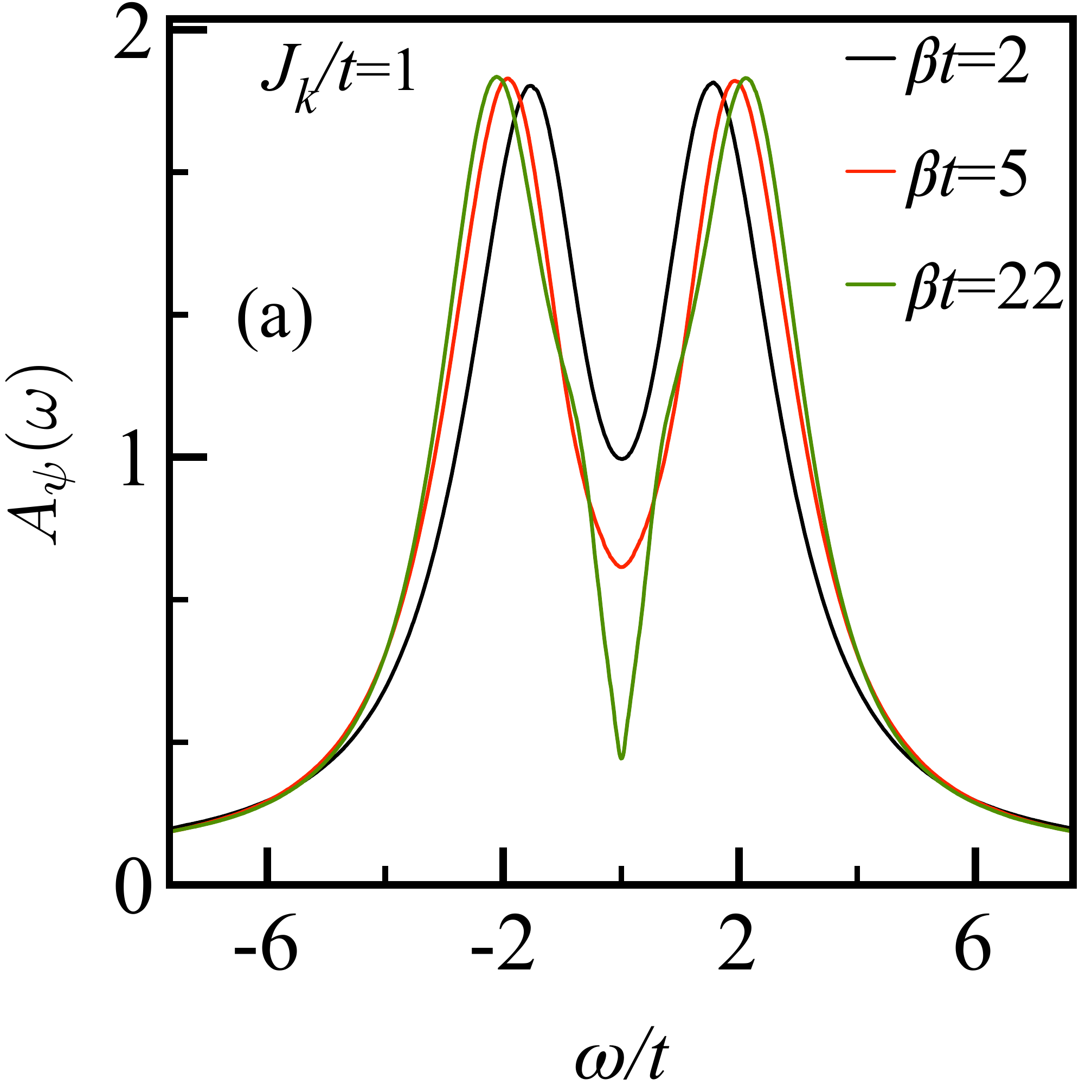}
\includegraphics[width=0.45\textwidth]{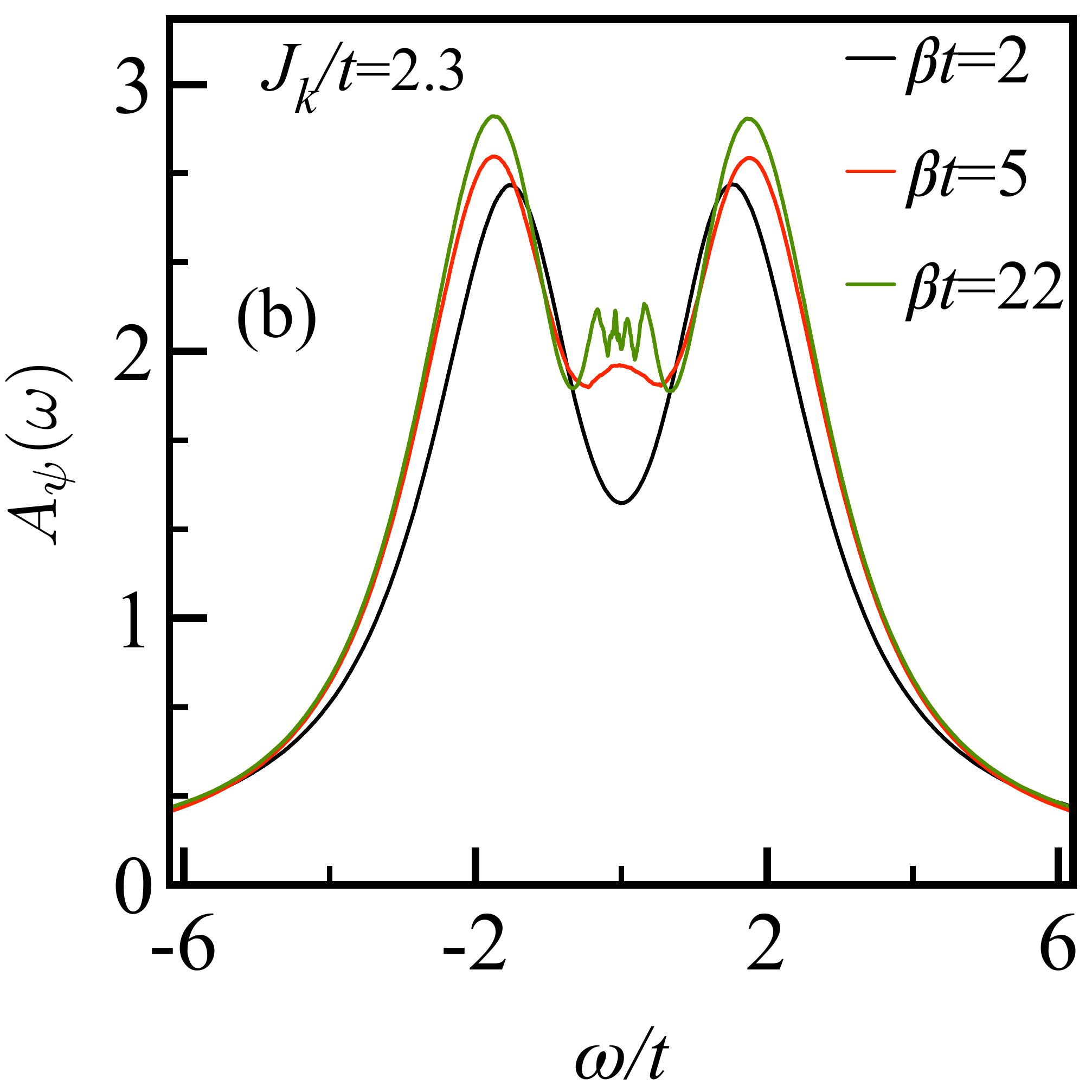}%
\caption{Local, i.e. momentum-integrated, composite-fermion spectral function $A_\psi(\omega)$  as a function of  energy $\omega/t$ at $L=L_x=L_y=44, J_h/t=1$. (a) In the ordered phase at $J_k/t=1$ and (b) in the paramagnetic phase at $J_k/t=2.3$.}
\label{Alpmega_vs_omega}
\end{figure}

Figs.~\ref{LogCr_Stau_vs_r_tau_betaLsqby4}   and \ref{LogCr_Stau_vs_r_tau_betaLsqby2}   plot  the  same  data as in    Fig.~4 of  the  main text  but  at
$\beta t  =  L^2/2$  and  $\beta t  =  L^2/4$.     For  $\beta t  =  L^2/4$   we  can  reach  larger  system sizes   and   the  same  overall  conclusions hold.
 Fig.~\ref{R_vs_1byL}   plots  the  correlation  ratio, $R$,  as  a  function of  system size    but  for  various aspect  ratios, $\beta t  =  L^2$,  $\beta  t =  L^2/2$  and  $\beta t  =  L^2/4$  as  well as as  a  function of  $J_k/t$.     Choosing  the  aspect ratio  $\beta  t =  L^2/4$  allows us  to  reach   larger  lattices.  As  apparent, the  data is  consistent   with   a slow increase  of  $R$    below  $J_k^{c}/t$. This  behavior is  characteristic of  the  crossover   regime,   where  the  lattice   sizes are  not  large enough  to  unambiguously   detect  long  range   antiferromagnetic ordering. Figs.~\ref{Alpmega_vs_omega} (a) and (b) plot the momentum-integrated composite fermion  spectral function $A_\psi(\omega)$  as a function of energy $\omega/t$ in ordered and disordered phases, respectively.

 \end{document}